\documentclass[prd, preprint, longbibliography, 11pt]{revtex4-1}

\usepackage{amsmath}
\usepackage{mathtools}
\usepackage{amssymb}
\usepackage{setspace}
\usepackage{graphicx}
\usepackage{natbib}
\usepackage{float}
\usepackage{tensor}
\usepackage[utf8]{inputenc}
\usepackage{amsfonts}
\usepackage{braket}
\usepackage{esint}

\usepackage{breqn}
\usepackage{IEEEtrantools}

\usepackage{blindtext}
\usepackage{graphicx}
\usepackage{amsmath}
\usepackage{dirtytalk}

\begin{document}
\begin{center}
{ \Large \bf   Majorana Neutrinos, Exceptional Jordan Algebra, and Mass Ratios for Charged Fermions}

\vskip 0.2 in

{\large{\bf Vivan Bhatt$^{a1}$, Rajrupa Mondal$^{b2}$, Vatsalya Vaibhav$^{c3}$ and Tejinder P.  Singh$^{d4}$ }}

\medskip

{\it $^a$Indian Institute of Technology Madras, 600036, India}\\

{\it $^c$Indian Institute of Science Education and Research, IISER Kolkata, 741246, India}\\

{\it $^c$Indian Institute of Technology Kanpur, 208016, India}\\

{\it $^d$Tata Institute of Fundamental Research, Homi Bhabha Road, Mumbai 400005, India}\\

\bigskip

 \;  {$^1$\tt vivanbhatt@smail.iitm.ac.in},   {$^{2}$\tt rm18ms099@iiserkol.ac.in}, {$^{3}$\tt vatsalya@iitk.ac.in},  \; {$^4$\tt tpsingh@tifr.res.in}

\bigskip



\end{center}


\centerline{\bf ABSTRACT}
   \noindent  We provide theoretical evidence that the neutrino is a Majorana fermion. This evidence comes from assuming that the standard model and beyond-standard-model physics can be described through division algebras, coupled to a quantum dynamics. We use the division algebras scheme to derive mass ratios for the standard model  charged fermions of three generations. The predicted ratios agree well with the observed values if the neutrino is assumed to be Majorana. However, the theoretically calculated ratios completely disagree with known values if the neutrino is taken to be a Dirac particle. Towards the end of the article we discuss prospects for unification of the standard model with gravitation if the assumed symmetry group of the theory is $E_6$, and if it is assumed that space-time is an 8D octonionic space-time, with 4D Minkowski space-time being an emergent approximation. Remarkably, we find evidence that the precursor of classical gravitation, described by the symmetry $SU(3)_{grav}\times SU(2)_R \times U(1)_{grav}$ is the right-handed counterpart of the standard model $SU(3)_{color}\times SU(2)_L \times U(1)_Y$. This provides the theoretical justification for the mass-ratios analysis based on the eigenvalues of the exceptional Jordan algebra.

\tableofcontents

\section{Introduction}
There exist a total of four normed division algebras in mathematics - the real numbers $\mathbb{R}$, the complex numbers $\mathbb{C}$, the quaternions $\mathbb{H}$ and the octonions $\mathbb{O}$. The first three of these algebras are used extensively in physics, albeit with the quaternions being used significantly lesser than the previous two. The primary hurdle for this was the non-commutativity of the quaternions and the non-associativity of the octonions. However, in recent years considerable research  has been done covering the intersection of division algebras, Clifford algebras and the standard model \cite{Baez1,Baez2,Baez3,Boyle1,Dixon1,Dray1,Todorov1,Furey1,Furey3,Furey2,Gillard1,Gunaydin1,Stoica1,Todorov2,Todorov3,Trayling2,Trayling3,Trayling1,Wilson1, chester}. Furey, in her thesis, obtained the first generation of fermions from the left ideals of Clifford algebras \cite{Furey1,Furey3,Furey2}. Related works are due to Gunaydin and Gursey \cite{Gunaydin1}, Stoica \cite{Stoica1}, Gresnigt \cite{Gillard1}, Wilson \cite{Wilson1} and Trayling and Baylis \cite{Trayling1}. In our previous work \cite{Tp3}, we had proposed using the second $SU(3)$ maximal subgroup from $F_4$ for generational symmetry; and we used the same octonionic representations to construct  exceptional Jordan matrices and used their eigenvalues to predict mass ratios.

Clifford algebras play a very important role in physics. They are an associative algebra generated by a vector space and a quadratic form. So we can generate a Clifford algebra from an underlying algebra of the vector space. Usually the vector space is taken to be matrix space for higher dimensional Clifford algebras but we can equally generate them using quaternions and octonions. The use of Clifford algebras in making fermions is based on the fact that we can make spinors from left ideals of Clifford algebras. The generating vectors of a complex Clifford algebra square to 1 and anti-commute with each other. For more details on Clifford algebras, please refer to \cite{Baez1,Clifford1,Furey1,Gallier1, Vatsalya1}.

Out of the five exceptional Lie Algebras discussed in mathematics \cite{Ramond1976, Yokota}, the underlying group of the lowest order among the five is $G_2$, which incidentally is the group of automorphisms of the octonions. The next in line is $F_4$ which is the group of automorphisms of the Exceptional Jordan Algebra. In Furey's thesis \cite{Furey1}, it has been shown that the fermionic states arise from simple octonionic chain algebra. Quite beautifully, the electric charge eigenvalues arise from the action of U(1) operator on those states. In our attempt to further look into what the rest of the exceptional Lie algebras reveal about the already known properties of the standard model, we investigate $F_4$. We then attempt to understand what the eigenvalues of $J_3(\mathbb{O})$ could possibly tell us  about the mass ratios of standard model fermions. Remarkably, this exercise gives us accurate results for the mass ratios \cite{Tp3, Vatsalya1} up to experimental error \cite{Zyla1}. In order to arrive at these mass ratios, we build upon our previous work \cite{Tp2,Tp4,Tp1,Tp3,Vatsalya1} and assume the neutrino to be a Majorana fermion. We then show that assuming the neutrino to be Dirac gives wrong mass ratios. Towards the end of the article we discuss prospects for unification of the standard model with gravitation if the assumed symmetry group of the theory is $E_6$, and if it is assumed that space-time is an 8D octonionic space-time, with 4D Minkowski space-time being an emergent approximation.

The application of the octonions to the standard model can be justified as follows. The standard model of particle physics has been a widely accepted theory for understanding the fundamental particles and the symmetries associated with their dynamics through the fundamental forces except gravity. The standard model gauge group is $G_{sm} = SU(3)_c\times SU(2)_L\times U(1)_Y$. The total of fifteen chiral fermions (considering only one generation) in the standard model can be represented under this gauge group as $(3, 2)_{Y_1}, (1, 2)_{Y_2}, (\overline{3}, 1)_{Y_3}, (\overline{3}, 1)_{Y_4}, (1, 1)_{Y_5}$ \cite{gg, slansky}. Interestingly these representations can be made using the octonions as well. The exceptional Lie group $G_2$ is the automorphism group of the octonions, i.e., it leaves the octonionic multiplication rule holomorphic. The octonionic multiplication rule is given by the following relation
\begin{equation}
    e_{\alpha}e_{\beta} = -\delta_{\alpha \beta} + g_{\alpha \beta \gamma}e_{\gamma}
\end{equation}
where the non trivial values of $g_{\alpha \beta \gamma}$ are given by $g_{235} = g_{346} = g_{615} = g_{672} = g_{574} = g_{371} = g_{124} = 1$. It is worth noting that unlike the quaternions, complex numbers and real numbers, the octonions are not associative. Therefore, they cannot be directly related to a Lie algebra but the automorphism group of the octonions $G_2$ is an exceptional Lie algebra. $G_2$ is a rank two Lie group therefore $SU(3)$ and $SU(2)\times SU(2)$ are the maximal subgroups of $G_2$. Let us write the purely imaginary octonions:
\begin{equation}
    \omega = \sum_{i=1}^{7}a_ie_i
\end{equation}
Here, the $a_i$s are real numbers and $e_i$s are the octonions. Using the multiplication rule of the octonions we can observe the following property
\begin{equation}
    e_{i+3} = e_7e_i
\end{equation}
Here the $e_i$s are quaternions and their automorphism group is SU(2). Therefore, we can write $\omega$ as:
\begin{equation}
    \omega = e_7 + \sum_{i=1}^{3}(a_i + e_7a_{3+i})e_i
\end{equation}
Therefore, $\omega$ which forms the fundamental representation of the $G_2$ group can be written as a sum of irreps of SU(3) using branching rules. This can be written as
\begin{equation}
    7 = 1 + 3 + \overline{3}
\end{equation}
If we rewrite $\omega$ in the following manner
\begin{equation}
    \omega = \sum_{i=1}^{3}a_ie_i + e_7(a_7 + a_4e_1 + a_5e_2 + a_6e_3)
\end{equation}
then the fundamental representation (7) of $G_2$ can be written as sum of irreps of $SU(2)\times SU(2)$ in the following manner:
\begin{equation}
    7 = (2,2) + (1,\overline{3})
\end{equation}
Therefore it is evident that using the octonions we can write all the chiral particle representations of the standard model. This is because the exceptional Lie group $G_2$ has $SU(3)$ and $SU(2)$ as its subgroup. In the next section we show the spinor representation of chiral fermions using the fact that spinors are left ideals of Clifford algebras.

\section{The complex Clifford algebra $\mathbb{C}l_6$ and unbroken $SU(3)\times U(1)$ for one generation of fermions}
We represent an octonion in the standard notation $(1, e_1, e_2, e_3, e_4, e_5, e_6, e_7)$ where the seven imaginary directions $e_i$ follow the Fano plane multiplication rules \cite{Vatsalya1}. 

It has been shown earlier that the Clifford algebra $\mathbb{C}l(6)$ made from complex octonionic chains can be used to describe the unbroken $SU(3)_c\times U(1)_{em}$ symmetry of one generation of standard model quarks and leptons, and their anti-particles.
The complex Clifford algebra $\mathbb{C}l(6)$ has a six dimensional generating vector space, the algebra is isomorphic to $C[8]$ which is the algebra of $8\times 8$ matrices with complex entries. Instead of working with $C[8]$ we can work with the complex octonionic chains $(\overleftarrow{\mathbb{C}}\otimes \overleftarrow{\mathbb{O}})$, which are defined to be maps acting on any element in $\mathbb{C}\otimes \mathbb{O}$ from left to right. Since all maps are associative, the octonionic algebra can be mapped to an associative algebra and therefore is isomorphic to the associative $\mathbb{C}l(6)$ algebra. 

We can  define the maximal totally isotropic subspace (MTIS) of the generating vector space of $\mathbb{C}l(6)$, and it is spanned by the following vectors \cite{Furey1}:
\begin{equation}
    \alpha_1 = \frac{-e_5 + ie_4}{2}, \quad \alpha_2 = \frac{-e_3 + i e_1}{2}, \quad \alpha_3 = \frac{-e_6 + ie_2}{2}
\end{equation}
The MTIS vectors obey the following commutation rules:
\begin{equation}
    \{\alpha_i, \alpha_j\} = 0, \quad \{\alpha_i^{\dagger}, \alpha_j^{\dagger}\} = 0, \quad \{\alpha_i, \alpha_j^{\dagger}\} = \delta_{ij}
\end{equation}
 Using this MTIS we can make spinors from the left-ideals of Clifford algebras which will be identified with  eight standard model fermions. If we define the idempotent as $\omega \omega^{\dagger} = \alpha_1\alpha_2\alpha_3\alpha_3^{\dagger}\alpha_2^{\dagger}\alpha_1^{\dagger}$ then the left action of $\mathbb{C}l(6)$ on this idempotent will give us one generation of fermions as follows:
\begin{align}
    V_{\overline{\nu}}&= \omega \omega^{\dagger} = \frac {1 + ie_{7}}{2}\qquad [\rm anti-Neutrino\ \rm singlet]\notag \\
\alpha_1^\dagger V_{\overline{\nu}}=\frac{e_5+ie_4}{2},\ \alpha_2^\dagger V_{\overline{\nu}}&=\frac{e_3+ie_1}{2},\ \alpha_3^\dagger V_{\overline{\nu}}=\frac{e_6+ie_2}{2} \ \qquad  [\rm Anti-Down\ \rm Quark\ \rm Triplet]\\
    \alpha_3^\dagger \alpha_2^\dagger V_{\overline{\nu}}=\frac{e_4+ie_5}{2},\ \alpha_1^\dagger\alpha_3^\dagger V_{\overline{\nu}}&=\frac{e_1+ie_3}{2},\ \alpha_2^\dagger\alpha_1^\dagger V_{\overline{\nu}}=\frac{e_2+ie_6}{2}\ \qquad [\rm Up\ \rm Quark\ \rm Triplet]\notag \\\
    \alpha_3^\dagger\alpha_2^\dagger\alpha_1^\dagger V_{\overline{\nu}}&=-\frac{i+e_7}{2}\ \qquad [\rm Positron\ singlet]\notag
 \label{fone}   
    \end{align}
Using the MTIS vectors we can write the following generator for $U(1)$ which provides electric charge to the fermions:
\begin{equation}
    Q = \frac{\alpha_1^{\dagger}\alpha_1 + \alpha_2^{\dagger}\alpha_2 + \alpha_3^{\dagger}\alpha_3}{3}
\end{equation}
The fermions shown as anti-down quarks are anti-triplets under $SU(3)$ and have a charge eigenvalue $Q=1/3$. Those labeled up quarks are triplets under $SU(3)$ and have a charge eigenvalue $Q=2/3$. The fermions labeled neutrino and positron are singlets under $SU(3)$ and respectively have charge eigenvalues $Q=0, 1$. This correct match between behaviour under $SU(3)$ and under $U(1)$ is highly non-trivial and justifies the particle identifications as shown against the states.
The automorphism group for the octonions is $G_2$ which has fourteen generators, eight of these generators can be used to generate the $SU(3)$ group. This $SU(3)_c$ group mediates color interaction amongst the quarks which come in three distinct colors. The $SU(3)$ generators are
\begin{align}
    \Lambda_1 &= -\alpha_2^{\dagger}\alpha_1 - \alpha_1^{\dagger}\alpha_2 & \Lambda_5 &= -i\alpha_1^{\dagger}\alpha_3 + i\alpha_3^{\dagger}\alpha_1 \\
    \Lambda_2 &= i\alpha_2^{\dagger}\alpha_1 - i\alpha_1^{\dagger}\alpha_2 & \Lambda_6 &= \alpha_3^{\dagger}\alpha_2 - \alpha_2^{\dagger}\alpha_3 \\
    \Lambda_3 &= \alpha_2^{\dagger}\alpha_2 - \alpha_1^{\dagger}\alpha_1 & \Lambda_7 &= i\alpha_3^{\dagger}\alpha_2 - i\alpha_2^{\dagger}\alpha_3 \\
    \Lambda_4 &= -\alpha_1^{\dagger}\alpha_3 - \alpha_3^{\dagger}\alpha_1 & \Lambda_8 &= -\frac{(\alpha_1^{\dagger}\alpha_1 + \alpha_2^{\dagger}\alpha_2 - 2\alpha_3^{\dagger}\alpha_3)}{\sqrt{3}}
\end{align}
The eight anti-particle states are obtained by first taking ordinary complex conjugation of the idempotent $V_{\overline{\nu}}$ which represented the anti-neutrino, i.e. $V^*_\nu = (1-ie_7)/2$ is the neutrino, and then by acting the MTIS generators on $V^*_\nu$. The $U(1)$  above is interpreted as $U(1)_{em}$. 

In the above analysis the neutrino is a left-handed Dirac neutrino \cite{Vatsalya1}. In the next section, we construct the eight fermion states if the neutrino were to be a Majorana neutrino, and then we construct the fermion states for the second and third generations, for both the Dirac neutrino case and the Majorana neutrino case. We will then use these states in the exceptional Jordan algebra to calculate mass-ratios of charged fermions and compare them with the experimentally observed mass-ratios.

\section{Octonionic Representations for Three Fermion Generations}
Closely following the interpretation of past authors \cite{Baez1,Baez2,Baez3,Furey1,Furey2,Furey3}, we have constructed the basis states of the minimal left ideal of $\mathbb{C}l(6)$ and identified one generation of leptons and quarks. Now we consider  two different sub-cases and further classify the neutrino as either a $\textbf{Dirac}$ or $\textbf{Majorana}$ fermion. For the Dirac case, the neutrino state retains the same expression as that of the left-handed Weyl spinor vaccum state worked out by Furey, $\textit{i.e.}$ \ $V_{\overline{\nu}}=\frac{1+ie_7}{2}$. This is because the Dirac neutrino can be written as a sum of right-handed Weyl spinor and left-handed Weyl spinor representation for the neutrino. Using the left-handed and right-handed values from our recent paper \cite{Vatsalya1}, the Dirac neutrino can be written as $V_D = (V_L+ V_R)/2 = (1+ie_7)/2$, whereas the Majorana neutrino can be written as $V_\nu^M = (V_{\overline{\nu}}-V^*_{\overline{\nu}})/2= ie_7/2$. We note that octonionic conjugation is denoted by a  tilde, complex conjugation by a *, and both together by a $\dagger$.

We now extend the analysis to the Majorana case. The Dirac neutrino can be written as $V_D = (V_L+ V_R)/2 = (1+ie_7)/2$, whereas the Majorana neutrino can be written as $V_\nu^M = (V_{\overline{\nu}}-V^*_{\overline{\nu}})/2= ie_7/2$. where $V_\nu^M $ is the Majorana neutrino, and as before $V_\nu$ is the Dirac neutrino. Armed with the vaccum states, we proceed to find the other states and hence the octonionic representations of the first generation of leptons and quarks. From the first generation, we propose that the subsequent generations are found by a method which we describe shortly.
\subsection*{Assuming Neutrino to be Majorana} Starting with the algebraic vacuum state as the Majorana neutrino, we get states for the first generation of quarks and leptons:
\begin{align}
    V_\nu^M&=\frac {ie_{7}}{2}\ \qquad [\rm Majorana\ \rm Neutrino]\notag \\
    \alpha_1^\dagger V_\nu^M=\frac{e_5+ie_4}{4},\ \alpha_2^\dagger V_\nu^M&=\frac{e_3+ie_1}{4},\ \alpha_3^\dagger V_\nu^M=\frac{e_6+ie_2}{4} \ \  \qquad [\rm Anti-Down\ \rm Quark\ \rm Triplet]\notag\\
    \alpha_3^\dagger \alpha_2^\dagger V_\nu^M=\frac{e_4+ie_5}{4},\ \alpha_1^\dagger\alpha_3^\dagger V_\nu^M&=\frac{e_1+ie_3}{4},\ \alpha_2^\dagger\alpha_1^\dagger V_\nu^M=\frac{e_2+ie_6}{4}\ \ \qquad[\rm Up\ \rm  Quark\ \rm Triplet]\notag \\
    \alpha_3^\dagger\alpha_2^\dagger\alpha_1^\dagger V_\nu^M&=-\frac{i+e_7}{4}\ \qquad [\rm Positron]\notag
\end{align}
Now, following the work done in \cite{Tp3}, we have obtained the representations of the first fermion generation. To use these as the elements of the Exceptional Jordan Algebra $J_3(\mathbb{O})$, we need to devise a map from the complex octonionic representation to a real octonionic one. Now, upon closer inspection, we can see that if color charge is specified (or ignored), then a set of Neutrino, Anti-down Quark, Up Quark and Positron has representation of a unique complex quaternionic subalgebra. For example, consider  $V_\nu^M,\ \alpha_1^\dagger V_\nu^M,\ \alpha_3^\dagger \alpha_2^\dagger V_\nu^M,\ \alpha_3^\dagger\alpha_2^\dagger\alpha_1^\dagger V_\nu^M$. It is clear that a quaternionic sub-group ($e_4,e_5,e_7$) completely represents the four states. Similarly for the other two colors, the quaternionic sub-groups ($e_1,e_3,e_7$) and ($e_2,e_6,e_7$) completely represent the four states.
Hence, for now, let us work with only the first sub-group $\textit{i.e.}$ ($e_4,e_5,e_7$). We will address this choice of the color charge in a later section and show that this choice has no effect on the results, as is only to be expected. We now take a quaternionic sub-representation
$$(a_0 + ia_1) + (a_2 + ia_3)e_4 + (a_4 + ia_5)e_5 + (a_6 + ia_7)e_7$$
where the  $a_i\in \mathbb{R}$.
We make the ansatz that this maps to the octonion
\begin{equation} a_0 + a_1e_1 + a_5e_2 + a_3e_3 + a_2e_4 + a_4e_5 + a_7e_6 + a_6e_7 \end{equation}
The four real coeffcients in the original complex quaternionic representation have been kept in
place, and their four imaginary counterparts have been moved to the octonion directions
($e_1, e_2, e_3, e_6$) respectively. The map thus constructed is invertible, $\textit{i.e.}$\,  given the real octonion, we can construct the equivalent complex quaternion representing the fermionic state.
Hence, under this mapping, the fermionic states for first generation, built from the Majorana neutrino,  are given as follows 
\begin{align}
    V_\nu^M=\frac {e_{6}}{2}, \quad
    V_{ad}=\frac{e_5+e_3}{4},\quad
    V_u=\frac{e_4+e_2}{4}\ \quad
    V_p=-\frac{e_1+e_7}{4}
\end{align}
Now, to arrive at the second and third generations of the fermions, we propose that the second generation states are obtained by a $\frac{2\pi}{3}$ rotation on the first generation state and the corresponding third generation state is obtained by a $\frac{2 \pi}{3}$ rotation on the second generation state. The motivation for this is that $F_4$, which is the automorphism group of $J_3(\mathbb{O})$ has $SU(3)\times SU(3)$ as one of its maximal subgroups. We already know that color appears out of one $SU(3)$ symmetry. We propose that the other $SU(3)$ symmetry gives rise to the three generations. SO(8) being the norm preserving group of octonions and there being 8 planes on an 8-dimensional sphere, each plane corresponds to a particle type with $2\pi/3$ rotations on each plane giving rise to a different generation. Also, since rotation matrices are unitary, we can use the simplest rotation operation to arrive at the second and third generations before coming back to the initial generation. For example, we propose that $V_{as}=e^{\frac {2 \pi e_5}{3}}V_{ad}$ and $V_{ab}=e^{\frac {2 \pi e_5}{3}}V_{as}=e^{\frac {4 \pi e_5}{3}}V_{ad}$. Now, here one might ponder as to why $e_5$ was specifically chosen as the imaginary unit as opposed to $e_3$. We will address and justify this seemingly ad hoc choice further in the paper, and show that we could as well have chosen $e_3$ without changing the results. For now, following the above hypothesis, we obtain the octonionic representations for the second and third generations of fermions, first assuming Majorana neutrino, then assuming Dirac neutrino.

\noindent \textbf{Generation II: (assuming Majorana neutrino)}
\begin{align}
    V_{\nu\mu}^M=-\frac {e_{6}+\sqrt{3}}{4}\qquad
    V_{as}=\frac{-e_5-e_3-\sqrt{3}-\sqrt{3}e_2}{8}\\
    V_c=\frac{-e_4-e_2-\sqrt{3}-\sqrt{3}e_1}{8}\qquad
    V_{a\mu}=\frac{e_1+e_7+\sqrt{3}-\sqrt{3}e_3}{8}\end{align}
\textbf{Generation III: (assuming Majorana neutrino)}
\begin{align}
    V_{\nu\tau}^M=-\frac {e_{6}-\sqrt{3}}{4}\qquad
    V_{ab}=\frac{-e_5-e_3+\sqrt{3}+\sqrt{3}e_2}{8}\\
    V_t=\frac{-e_4-e_2+\sqrt{3}+\sqrt{3}e_1}{8}\qquad
    V_{a\tau}=\frac{e_1+e_7-\sqrt{3}+\sqrt{3}e_3}{8}\notag
\end{align}
\subsection*{Assuming a Dirac neutrino}
The same calculations and theoretical arguments as above are repeated for the Dirac neutrino case. Before mapping the complex octonions to the real ones, the states obtained are same as earlier above:
\begin{align}
    V_{{\overline{\nu}}}&= \omega \omega^{\dagger} = \frac {1 + ie_{7}}{2}\qquad [\rm  Dirac\ Neutrino\ \rm singlet]\notag \\
\alpha_1^\dagger V_{\overline{\nu}}=\frac{e_5+ie_4}{2},\ \alpha_2^\dagger V_{\overline{\nu}}&=\frac{e_3+ie_1}{2},\ \alpha_3^\dagger V_{\overline{\nu}}=\frac{e_6+ie_2}{2} \ \qquad  [\rm Anti-Down\ \rm Quark\ \rm Triplet]\\
    \alpha_3^\dagger \alpha_2^\dagger V_{\overline{\nu}}=\frac{e_4+ie_5}{2},\ \alpha_1^\dagger\alpha_3^\dagger V_{\overline{\nu}}&=\frac{e_1+ie_3}{2},\ \alpha_2^\dagger\alpha_1^\dagger V_{\overline{\nu}}=\frac{e_2+ie_6}{2}\ \qquad [\rm Up\ \rm Quark\ \rm Triplet]\notag \\\
    \alpha_3^\dagger\alpha_2^\dagger\alpha_1^\dagger V_{\overline{\nu}}&=-\frac{i+e_7}{2}\ \qquad [\rm Positron\ singlet]\notag
\end{align}
After following the same mapping as in the case of the Majorana neutrino consideration, the three generation octonionic states are as follows.

\textbf{Generation I: (assuming Dirac neutrino)}
\begin{align}
    V_{\overline{\nu}}=\frac {1+e_{6}}{2}\qquad
    V_{ad}=\frac{e_5+e_3}{2}\qquad
    V_u=\frac{e_4+e_2}{2}\qquad
    V_p=-\frac{e_1+e_7}{2}
\end{align}
\textbf{Generation II: (assuming Dirac neutrino)}
\begin{align}
    V_{\mu\overline{\nu}}=-\frac {e_{6}+\sqrt{3}}{2}\qquad
    V_{as}=\frac{-e_5-e_3-\sqrt{3}-\sqrt{3}e_2}{4}\\
    V_c=\frac{-e_4-e_2-\sqrt{3}-\sqrt{3}e_1}{4}\qquad
    V_{a\mu}=\frac{e_1+e_7+\sqrt{3}-\sqrt{3}e_3}{4}
\end{align}
\textbf{Generation III (assuming Dirac neutrino)}:
\begin{align}
    V_{\tau\overline{\nu}}=-\frac {e_{6}-\sqrt{3}}{2}\qquad
    V_{ab}=\frac{-e_5-e_3+\sqrt{3}+\sqrt{3}e_2}{4}\\
    V_t=\frac{-e_4-e_2+\sqrt{3}+\sqrt{3}e_1}{4}\qquad
    V_{a\tau}=\frac{e_1+e_7-\sqrt{3}+\sqrt{3}e_3}{4}\notag
\end{align}
\section{Exceptional Jordan Matrices by Family, and their eigenvalues}
Above we have found two sets of octonionic representations for three generations of standard model fermions; one set assuming the neutrino to be a Dirac fermion, and the other assuming the neutrino to be Majorana. We will now use these representations in the exceptional Jordan algebra, to find the eigenvalues of its characteristic equation. These eigenvalues will then be justified to be square-root mass numbers, which will hence be used to find mass ratios.

A general matrix of the Exceptional Jordan Algebra $J_3(\mathbb{O})$ can be written as:
\begin{align}
    X(\xi,x) &= \begin{bmatrix}
                \xi_1 & x_3 & \tilde{x}_2\\
                \tilde{x}_3 & \xi_2 & x_1\\ 
                x_2 & \tilde{x}_1 & \xi_3
            \end{bmatrix}
\end{align}
its characteristic equation is a cubic:
\begin{equation}X^3-Tr(X)X^2+S(X)X-Det(x)=0 \end{equation}
where
\begin{equation}Tr(x)=\xi_1+\xi_2+\xi_3\ ,\ \ \ \  Det(X)=\xi_1 \xi_2 \xi_3 +2\ Re(x_1 x_2 x_3)-\sum\limits_{i=1}^3 \xi_i x_i \tilde{x}_i\end{equation}
\begin{equation}S(x)=\xi_1 \xi_2+\xi_2 \xi_3+\xi_3 \xi_1-x_1\tilde{x}_1-\tilde{x}_2x_2-x_3\tilde{x}_3 \end{equation}
Here, the $'\xi's \in \mathbb{R}$ and the octonions are defined on $\mathbb{R}$.
We  propose, and will justify it further in the subsequent sections,  that the roots of this equation give information about the experimentally known mass ratios of quarks and leptons.
Closely following Baez's argument that projections of EJA to $\mathbb{O}P^2$ takes one of the following four forms (upto automorphisms):
\begin{align}
    p_0  = \begin{bmatrix}
                0 & 0 & 0\\
                0 & 0 & 0\\ 
                0 & 0 & 0
            \end{bmatrix}
            \quad
    p_1 =  \begin{bmatrix}
                1 & 0 & 0 \\
                0 & 0 & 0\\ 
                0 & 0 & 0
            \end{bmatrix}
            \quad
    p_2  = \begin{bmatrix}
                1 & 0 & 0 \\
                0 & 1 & 0 \\ 
                0 & 0 & 0
            \end{bmatrix}
            \quad
    p_3 =  \begin{bmatrix}
                1 & 0 & 0 \\
                0 & 1 & 0 \\ 
                0 & 0 & 1
            \end{bmatrix}
\end{align}
We see that the invariant traces are the eigenvalues of the number operator defined by Furey, which gives rise to $3\times$ [charge of the $U(1)$ generator] in the $\mathbb{C}l(6)$ left ideal. We thus originally proposed to identify the trace with the sum of the charges of the three identically charged fermions, and the individual diagonal entries as the electric charge. As discussed above, here we are working with the $SU(3)_{Generations}$. And as per the observed mass ratios, the positron : up quark : down quark square-root mass ratios are $1:2:3$. So, in our case, the square-root mass number (gravi-charge) of the first generation of charged fermions will be $\frac{1}{3},\frac{2}{3},1$ for positron, up quark and anti-down quark respectively. Hence, the trace 0 $J_3(\mathbb{O})$ matrix will represent three generations of neutrinos, trace 1 $J_3(\mathbb{O})$ matrix the three generations of electrons, trace 2 the three up quark generations while trace 3 the three down quark ones. It is important to note that on a comparison between the gravi-charge and electric charge, the relative position of the up quark remains the same while that of the down quark and the electron interchanges. More on this will be discussed in a later section.

Thus, the corresponding Jordan matrices can be written in the following way:
\begin{align}
    X_\nu &= \begin{bmatrix}
                0 & V_{\tau} & \tilde{V_{\mu}}\\
                \tilde{V_{\tau}} & 0 & V_{\nu}\\ 
                V_{\mu} & \tilde{V_{\nu}} & 0
            \end{bmatrix}\quad
        X_{e} &=  \begin{bmatrix}
                \frac{1}{3} & V_{a\tau} & \tilde{V_{a\mu}} \\
                \tilde{V_{a\tau}} & \frac{1}{3} & V_{e^+} \\ 
                V_{a\mu} & \tilde{V_{e^+}} & \frac{1}{3}
            \end{bmatrix}\quad
    X_{u} &=\begin{bmatrix}
                \frac{2}{3} & V_{t} & \tilde{V_{c}} \\
                \tilde{V_{t}} & \frac{2}{3} & V_{u} \\ 
                V_{c} & \tilde{V_{u}} & \frac{2}{3}
            \end{bmatrix}\quad
    X_{d} &= \begin{bmatrix}
                1 & V_{ab} & \tilde{V_{as}} \\
                \tilde{V_{ab}} & 1 & V_{ad} \\ 
                V_{as} & \tilde{V_{ad}} & 1\\
            \end{bmatrix}
\end{align}
Using the octonionic representations for the fermions, as constructed above, in these matrices, we now find the roots of the cubic characteristic equation, for the various cases:

\noindent {\it States made from Majorana Neutrino:}

\noindent For the \textbf{Majorana neutrino}, the cubic equation and its roots are 
\begin{equation}
Tr(X_\nu)=0,\ S(X_\nu)=-\frac{3}{4},\ Det(X_\nu)=0, \qquad
x^3-\frac{3}{4}x=0, \qquad
\boxed{-\frac{\sqrt{3}}{2}\ \ 0\ , \frac{\sqrt{3}}{2}}
\end{equation}
For the \textbf{positron},
\begin{equation}
Tr(X_e)=1,\ S(X_e)=-\frac{7}{6},\ Det(X_e)=-\frac{25}{54},\qquad
x^3-x^2-\frac{7}{6}x+\frac{25}{54}=0,\qquad
\boxed{\frac{1}{3}-\sqrt{\frac{3}{8}}\ ,\ \frac{1}{3}\ , \frac{1}{3}+\sqrt{\frac{3}{8}}}
\end{equation}
For the \textbf{up quark},
\begin{equation}
Tr(X_u)=2,\ S(X_u)=\frac{23}{24},\ Det(X_u)=\frac{5}{108},\qquad
x^3-2x^2+\frac{23}{24}x-\frac{5}{108}=0, \qquad
\boxed{\frac{2}{3}-\sqrt{\frac{3}{8}}\ ,\ \frac{2}{3}\ , \frac{2}{3}+\sqrt{\frac{3}{8}}}
\end{equation}
For the \textbf{antidown quark},
\begin{equation}
Tr(X_d)=3,\ S(X_d)=-\frac{1}{24},\ Det(X_d)=-\frac{19}{216}\qquad
x^3-3x^2-\frac{1}{24}x+\frac{19}{216}=0\qquad
\boxed{1-\sqrt{\frac{3}{8}}\ ,\ 1\ ,\ 1+\sqrt{\frac{3}{8}}}
\end{equation}

\noindent{\it States made from the Dirac Neutrino:}

\noindent For the \textbf{Dirac neutrino},
\begin{equation}
Tr(X_\nu)=0,\ S(X_\nu)=-\frac{3}{2},\ Det(X_\nu)=-\frac{1}{2}, \qquad
x^3-\frac{3}{2}x+\frac{1}{2}=0,\qquad
\boxed{-\frac{1}{2}-\frac{\sqrt{3}}{2}\ ,\ 1\ , -\frac{1}{2}+\frac{\sqrt{3}}{2}}
\end{equation}
For the \textbf{positron},
\begin{equation}
Tr(X_e)=1,\ S(X_e)=-\frac{7}{6},\ Det(X_e)=-\frac{25}{54},\qquad
x^3-x^2-\frac{7}{6}x+\frac{25}{54}=0,\qquad
\boxed{\frac{1}{3}-\sqrt{\frac{3}{2}}\ ,\ \frac{1}{3}\ , \frac{1}{3}+\sqrt{\frac{3}{2}}}
\end{equation}
For the \textbf{up quark},
\begin{equation}
Tr(X_u)=2,\ S(X_u)=-\frac{1}{6},\ Det(X_u)=-\frac{179}{216}, \qquad
x^3-2x^2-\frac{1}{6}x+\frac{179}{216}=0, \qquad
\boxed{\frac{2}{3}-\sqrt{\frac{3}{2}}\ ,\ \frac{2}{3}\ , \frac{2}{3}+\sqrt{\frac{3}{2}}}
\end{equation}
For the \textbf{antidown quark},\qquad
\begin{equation}
Tr(X_d)=3,\ S(X_d)=\frac{3}{2},\ Det(X_d)=-\frac{1}{2}\qquad
x^3-3x^2+\frac{3}{2}x+\frac{1}{2}=0\qquad
\boxed{1-\sqrt{\frac{3}{2}}\ ,\ 1\ , 1+\sqrt{\frac{3}{2}}}
\end{equation}
We observe that as we go from the Majorana neutrino case to the Dirac neutrino case, the roots for the neutrinos change significantly, and in the roots for the charged fermions, the factor of $\sqrt{3/8}$ gets replaced everywhere by $\sqrt{3/2}$. This makes a crucial difference to the mass ratios as we will see, with the Dirac neutrino leading to ratios which do not agree with known values.

\section{Exceptional Jordan Matrices by Generation}
The primary aim of this paper is to study if we can construct the fermionic mass ratios from first principles of the Exceptional Jordan Algebra. As per the eigenvalues calculated in the previous section, it is seen that the two eigenvalues that do not correspond to the electric charge are shifted symmetrically around the middle eigenvalue which is equal to the electric charge \cite{Tp3}. This gives zero to be one of the eigenvalues for the neutrino family, which seems to suggest that we cannot obtain the neutrino masses from these constructions, and that neutrino masses would thus arise from some alternate mechanism. We discuss this further in Sections 8 and 9. 

For the time being, we focus on the charged fermions and check if any other eigenvalues calculated  from $J_3(\mathbb{O})$ give us the desired mass ratios. We begin by constructing the Jordan matrices by generation, as opposed to those constructed by family in the previous section. This gives us the following three matrices with real octonionic entries:
\begin{align}
    X_I =  \begin{bmatrix}
                1 & V_{e^+} & V_{up}^* \\
                V_{e^+}^* & \frac{2}{3} & V_{ad} \\ V_{up} & V_{ad}^* & \frac{1}{3}
            \end{bmatrix}\qquad
    X_{II} =  \begin{bmatrix}
                1 & V_{a\mu} & V_{c}^* \\
                V_{a\mu}^* & \frac{2}{3} & V_{as} \\ V_{c} & V_{as}^* & \frac{1}{3}
            \end{bmatrix}\qquad
    X_{III} =  \begin{bmatrix}
                1 & V_{a\tau} & V_{t}^* \\
                V_{a\tau}^* & \frac{2}{3} & V_{ab} \\ V_{t} & V_{ab}^* & \frac{1}{3}
            \end{bmatrix}
\end{align}
We use the same notation and octonionic representations used earlier in the paper and note that $Tr(x)$ is the same across the generations and equal to two. Further, we find that $S(X)$ is also invariant for both Dirac and Majorana cases. $Det(X)$, however, changes with the generation due to different values of $Re(x_1x_2x_3)$.  Hence, $Tr(X) = 2, 
    S(X) = \frac{61}{72}$.  We then solve the characteristic equation to get nine unequal eigenvalues. Henceforth, we refer to these eigenvalues as the {vertical} eigenvalues, since they are calculated by generation and not by family -  these family ones  will be called {horizontal} eigenvalues. 

{\it Majorana Neutrino set:}
Taking the octonionic representations corresponding to a Majorana neutrino, we calculate the determinants by generation
\begin{equation}
    Det(X_I) = \frac{-25-9}{576}, \hspace{0.5cm} Det(X_{II}) = \frac{-25+\sqrt{3}}{576}, \hspace{0.5cm}  Det(X_{III}) = \frac{-25-\sqrt{3}}{576}
\end{equation}
\noindent Thus, we find the vertical eigenvalues for the Majorana case. 
\begin{equation}
    \begin{matrix}
        \lambda_{I_1}= -0.04318 & \lambda_{I_2}= 0.69266 & \lambda_{I_3}= 1.35052 \\
        \lambda_{II_1}= -0.04898 & \lambda_{II_2}= 0.70511 & \lambda_{II_3}= 1.34387 \\
        \lambda_{III_1}= -0.06071 & \lambda_{III_2}= 0.73151 & \lambda_{III_3}= 1.32919
    \end{matrix}
\end{equation}
{\it Dirac Neutrino set:}
Next, we use the Dirac neutrino based representation to calculate the following vertical eigenvalues:
\begin{equation}
    \begin{matrix}
        \lambda_{I_1}= 1.8498 & \lambda_{I_2}= -0.67407 & \lambda_{I_3}= 0.82427 \\
        \lambda_{II_1}= 1.85622 & \lambda_{II_2}= -0.66962 & \lambda_{II_3}= 0.81341 \\
        \lambda_{III_1}= 1.83842 & \lambda_{III_2}= -0.68168 & \lambda_{III_3}= 0.84326
    \end{matrix}
\end{equation}
We will return to an analysis of these eigenvalues later in the paper.

\section{Checking for the invariance of the eigenvalues}
In the preceding two sections, we calculate a total of 42 eigenvalues - 21 for the Dirac case and 21 for the Majorana. Each of these 21 eigenvalues are further divided into 12 horizontal and 9 vertical eigenvalues. Keeping the neutrino aside, we are still left with 36 eigenvalues. Constructing 6 mass ratios by operating on 36 different numbers would not be a very difficult or noteworthy task, purely by means of the mathematical freedom available to us. However, we will methodically show that out of these 36 eigenvalues, only 9 hold the key to the fermionic mass ratios. We address the Majorana neutrino v/s Dirac neutrino question in a later section, and currently turn our attention towards the horizontal and vertical eigenvalues.

If our above-calculated eigenvalues are indeed to have relevance to the standard model, they should survive further checks on their invariant nature, which we now apply.

\noindent {\it Invariance under change of color charge:} 

In Section 2, we obtained two triplets from the Clifford algebra that Furey identified with the up and antidown quarks \cite{Furey1}. To calculate the eigenvalues, we had to map our octonions from $\mathbb{C}\times\mathbb{O}$ to $\mathbb{R}\times\mathbb{O}$, and we chose a particular pair of up and antidown quark to proceed with the same. In so doing, we temporarily set aside the $SU(3)$ associated with color, which physically corresponds to choosing a particular color charge and finding the eigenvalues for said color charge. If our eigenvalues are related to the fermionic mass ratios, they need to be the same for all three color charges even if the individual Jordan matrices use different representations. 

 We originally worked with $V_{ad_1}$ and $V_{u_1}$, but we now solve the eigenvalue problem for $V_{ad_2}$ and $V_{u_2}$. Thus, the four first generation fermionic representations we deal with (for the Majorana case) are
\begin{align}
    V_{{\nu}} = \frac{ie_7}{2}\qquad
    V_{ad} =\notag \frac{e_3+ie_1}{4}\qquad
    V_u = \frac{e_1+ie_3}{4}\qquad
    V_{e^+} = -\frac{i+e_7}{4}
\end{align}
Again, we map these representations from $\mathbb{C}\times\mathbb{O}$ to $\mathbb{R}\times\mathbb{O}$. Our quaternionic sub-representation
\begin{equation} (a_0 + ia_1) + (a_2 + ia_3)e_1 + (a_4 + ia_5)e_3 + (a_6 + ia_7)e_7\end{equation}
where the 'a's $\in \mathbb{R}$,
maps to 
\begin{equation} a_0 + a_2e_1 + a_1e_2 + a_4e_3 + a_3e_4 + a_5e_5 + a_7e_6 + a_6e_7\end{equation}  
without loss of generality (any other map also leads us to the same pattern in our results). Using the above representations, we determine all the eigenvalues again and some tedious but straightforward calculations show that the horizontal eigenvalues remain the same after changing the color, as the trace, determinant and $S(X)$ remain constant. However, for the vertical eigenvalues, the product $x_1x_2x_3$ changes for different colors, which, in turn, changes the determinants and thus the eigenvalues. The vertical eigenvalues appear to have no invariance, as further checks also show, and hence they have no evident physical significance. Only the 12 horizontal eigenvalues are significant, and leaving out the three for the neutrinos, the remaining nine determine the mass ratios for the nine charged fermions.

\noindent {\it Charge conjugation:}
In Furey's work, antiparticles are represented as the complex conjugate of their corresponding particles. Following the same construction here, it is imperative that the eigenvalues are invariant to complex conjugation of the octonions in $\mathbb{C}\times\mathbb{O}$ before mapping them to $\mathbb{R}\times\mathbb{O}$. Only in this case can we hope to relate them to the mass ratios as the particle-antiparticle pairs differ only by a change in sign of the charge and not the mass. Of course, we alter the charge accordingly to obtain the following Jordan matrices:

 \textbf{Horizontal Matrices:}
\begin{align}
    X_\nu = \begin{bmatrix}
                0 & V_{a\tau} & \tilde{V_{a\mu}}\\
                \tilde{V_{a\tau}} & 0 & V_{a\nu}\\ 
                V_{a\mu} & \tilde{V_{a\nu}} & 0
            \end{bmatrix}\qquad      
    X_{e^-} =  \begin{bmatrix}
                -\frac{1}{3} & V_{\tau} & \tilde{V_{\mu}} \\
               \tilde{ V_{\tau}} & -\frac{1}{3} & V_{e^-} \\ 
                V_{\mu} & \tilde{V_{e^-}} & -\frac{1}{3}
            \end{bmatrix}\qquad
    X_{u} =\begin{bmatrix}
                -\frac{2}{3} & V_{at} & \tilde{V_{ac}} \\
                \tilde{V_{at}} & -\frac{2}{3} & V_{au} \\ 
                V_{ac} & \tilde{V_{au}} & -\frac{2}{3}
            \end{bmatrix}\qquad
    X_{d} = \begin{bmatrix}
                -1 & V_{b} & \tilde{V_{s}} \\
                \tilde{V_{b}} & -1 & V_{d} \\ 
                V_{s} & \tilde{V_{d}} & -1\\
            \end{bmatrix}
           \end{align}

 \textbf{Vertical Matrices:}
\begin{align}
    X_I =  \begin{bmatrix}
                -1 & V_{e^-} & \tilde{V_{aup}} \\
                \tilde{V_{e^-}} & -\frac{2}{3} & V_{d} \\ V_{aup} & \tilde{V_{d}} & -\frac{1}{3}
            \end{bmatrix}\qquad
    X_{II} =  \begin{bmatrix}
                -1 & V_{\mu} & \tilde{V_{ac}} \\
                \tilde{V_{\mu}} & -\frac{2}{3} & V_{s} \\ V_{ac} & \tilde{V_{s}} & -\frac{1}{3}
            \end{bmatrix}\qquad
    X_{III} =  \begin{bmatrix}
                -1 & V_{\tau} & \tilde{V_{at}} \\
                \tilde{V_{\tau}} & -\frac{2}{3} & V_{b} \\ V_{at} & \tilde{V_{b}} & -\frac{1}{3}
            \end{bmatrix}
\end{align}
Interestingly enough, we again find that the horizontal eigenvalues remain invariant up to sign, whereas the vertical eigenvalues are completely different. The fact that the horizontal eigenvalues change sign when the sign of electric charge is changed, encourages us to associate a square-root mass number $\pm\sqrt{m/m_{Pl}}$ with standard model fermions, given by the horizontal eigenvalues. It is noteworthy then that a particle and its anti-particle will have the same value for $e\sqrt{m}$; this quantity does not change sign under charge conjugation.

\noindent {\it Generational $SU(3)$: }
Now, we address our last assumption while choosing the octonionic axes of rotations for the $SU(3)_{Generation}$. In section 3, we put forth our suggestion that the last remaining $SU(3)$ gave rise to the three fermionic generations, and used rotations about octonionic axes to obtain the representations for the second and third generations of fermions. This choice of axis, however, was not unique and was chosen by the authors to enable the explicit representation required to solve the eigenvalue problem. It thus becomes necessary to ensure that this arbitrary choice did not affect the eigenvalues as that goes against the philosophy of both this paper and the whole division algebra approach where we suggest that the free parameters of the Standard Model need not be put in by hand and actually emerge from the octonionic algebra.

For example, we rotated $V_{ad}$ by $\frac{2\pi}{3}$ by left multiplying it by $e^{\frac{2\pi e_5}{3}}$. Since $e_5$ and $e_3$ are both equivalent here, we could have chosen $e_3$ as well. We carry out this exercise and find new octonionic representations for the two higher generations
\begin{align}
    V_{as} = \frac{-e_5-e_3+\sqrt{3}e_2-\sqrt{3}}{8}\qquad 
    V_{ab} = \frac{-e_5-e_3-\sqrt{3}e_2+\sqrt{3}}{8}
\end{align}
Using these and other similar representations different from the original ones, we repeat our entire process and find that both sets of horizontal eigenvalues do not change, whereas their vertical counterparts do change.

\noindent {\bf \it Horizontal eigenvalues are invariants:} We had made three seemingly ad hoc choices in the process of calculating the eigenvalues, and have now studied the consequences of all three choices. We varied the color charge, found the eigenvalues for both particles and antiparticles, and tried different octonionic rotations as well. For all of these, the horizontal eigenvalues remained unchanged whereas their vertical counterparts were different in all these cases. The above exercise might make the process of finding the vertical eigenvalues seem to be a futile effort. Although we do not pursue the vertical eigenvalues further at this stage and state that they are not related to the fermionic mass ratios, the generational calculations are not in vain. 

The fact that the vertical eigenvalues do, in fact, change, shows that the invariance of the horizontal eigenvalues is no mere mathematical coincidence. This non-trivial invariance strongly suggests that the horizontal eigenvalues do have a special place in $J_3(\mathbb{O})$, which has previously been shown to be strongly related to the standard model as we know it today. It is not a leap of faith, then, to suggest that these very eigenvalues would give rise to other hitherto unexplored parameters of the standard model. In the coming sections, we show that the observed fermionic mass ratios emerge from these horizontal eigenvalues, in a method parallel to Furey's derivation of quantized charge \cite{Furey1}. 

The eigenmatrices corresponding to the horizontal eigenvalues can also be worked out, both for the Majorana neutrino case, and for the Dirac neutrino case. These results are presented in an Appendix at the end of the paper.

\section{Fermionic Mass Ratios}
In this section we calculate  the square-root of mass ratios of the charged fermions  using the eigen-values of the exceptional Jordan algebra. For a particular family of fermions the product of these eigen-values gives us the square root mass ratio of second and third generation fermion with respect to the first generation fermion. The theoretical values are then compared with the experimentally known square-root mass ratios  obtained by dividing one known mass with respect to the other, and taking the square root. Measured mass values have been taken from \cite{Zyla1}.  Why do the Jordan eigenvalues, which we identify as square-root mass numbers, tell us about mass ratios? We return to this question in some detail in Section X, where in the context of the exceptional group $E_6$ we propose, with justification,  that one of the $SU(3)$ sub-groups of $E_6$ be identified with gravitational charge (equivalently, square-root mass number), analogous to the $SU(3)_c$ for QCD color. For now, we can justify the analysis below by saying that the charge eigenstates obtained from the octonion algebra are not mass eigenstates, but rather, superpositions of the latter. The eigen-matrices are mass eigenstates, and the corresponding eigenvalues are square-root mass numbers. For reasons that become clearer in Section IX, the Jordan eigenvalues of the down quark family and electron family are interchanged for the purpose of calculating theoretical mass ratios. Put briefly, this becomes necessary because the down : up : electron square-root mass ratios 3 : 2 : 1 are in reverse order to their electric charge ratios 1 : 2 : 3

\subsection*{Root-mass ratios for Majorana neutrino set}  For ease of reference, the Jordan eigenvalues used for finding square-root mass ratios are summarised in Fig. 1 below.
\begin{figure}[h]
\centering
\includegraphics[width=10.5cm]{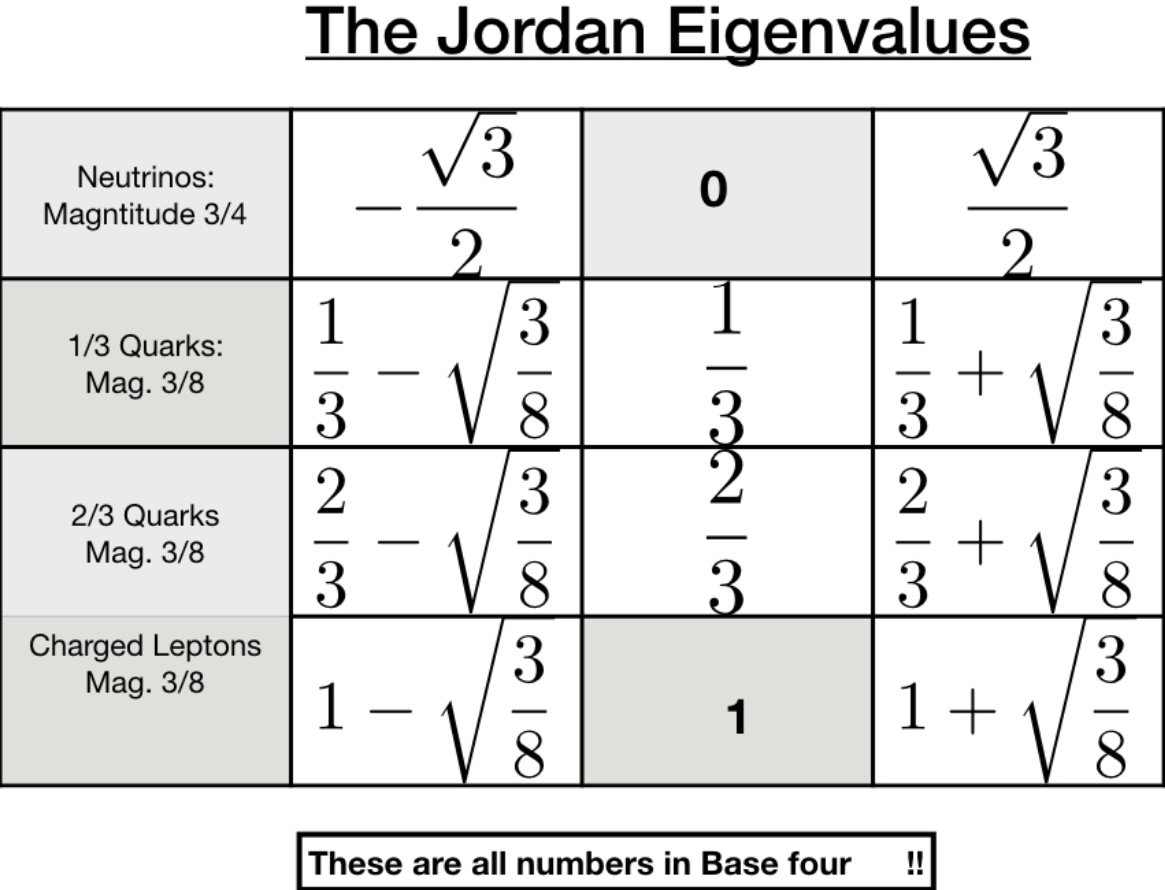}
\caption{The eigenvalues of exceptional Jordan matrices for various fermions, assuming neutrino to be Majorana \cite{Tp3}}
\end{figure}
 The square-root mass ratios of fermions, assuming the neutrino to be Majorana, are now calculated.
\begin{itemize}
\item Strange quark with respect to down quark: 
\begin{equation}
    \frac{1+\sqrt{3/8}}{1-\sqrt{3/8}} = 4.16; \quad \sqrt{\frac{95}{4.7}} = 4.50
    \end{equation}
    The charge 1 eigenvalues are assigned to the down quark family, with the largest value given to strange quark, and smallest to down quark. The theoretical prediction 4.16 lies jut outside the experimental range (4.21, 4.86) of the corresponding ratio (see Table I below). 
\item Bottom quark with respect to down quark:
\begin{equation}
    \frac{1+\sqrt{3/8}}{1-\sqrt{3/8}}\times\frac{1+\sqrt{3/8}}{1}\times\frac{1+\sqrt{3/8}}{1-\sqrt{3/8}} = 28.44; \quad \sqrt{\frac{4180}{4.7}} = 29.82
\end{equation}
The strange to down ratio has been squared, and multiplied by the largest eigenvalue. The theoretical prediction 28.44 lies within the experimentally measured range (28.25, 30.97).
\item Charm quark with respect to up quark
\begin{equation}
    \frac{2/3+\sqrt{3/8}}{2/3-\sqrt{3/8}} = 23.57; \quad \sqrt{\frac{1275}{2.3}} = 23.55
\end{equation}
The largest eigenvalue is divided by the smallest eigenvalue, and the theoretical prediction of 23.57 lies within the experimental range (21.04, 26.87). 
\item Top quark with respect to up quark
\begin{equation}
    \frac{2/3+\sqrt{3/8}}{2/3-\sqrt{3/8}}\times\frac{2/3}{2/3-\sqrt{3/8}} = 289.26; \quad \sqrt{\frac{173210}{2.3}} = 274.42
\end{equation}
The charm to up ratio is multiplied by the ratio of the middle to the smallest eigenvalue. The theoretical value of 289.26 lies within the experimental range (248.18, 310.07). How do these small fractions manage to generate the huge top quark mass?! The answer lies in the numerical coincidence that $2/3\approx 0.67$ is very close to $\sqrt{3/8}\approx 0.61$ so that $(2/3-\sqrt{3/8})^{-2}\approx 339$ gives a gain factor of over 300, making the top quark so heavy. We take this numerical coincidence as a serious indicator that this theory is on the right track. For we will see shortly, that when the Dirac neutrino is assumed, the $\sqrt{3/8}$ is replaced by $\sqrt{3/2}$ and the theoretical prediction for the top to up ratio goes completely wrong.
\item Muon with respect to electron:
\begin{equation}
    \frac{1+\sqrt{3/8}}{1-\sqrt{3/8}}\times\frac{1/3+\sqrt{3/8}}{|1/3-\sqrt{3/8}|} = 14.10; \quad \sqrt{206.7682830} = 14.38
    \end{equation}
    The ratio of the largest to smallest eigenvalue for the electron family has been multiplied by the strange to down ratio. There is about $4\%$ deviation from the known mass ratio of the muon to the electron.
\item Tau-lepton with respect to electron
\begin{equation}
    \frac{1+\sqrt{3/8}}{1-\sqrt{3/8}}\times\frac{1/3+\sqrt{3/8}}{|1/3-\sqrt{3/8}|}\times\frac{1+\sqrt{3/8}}{1-\sqrt{3/8}} = 58.64; \quad \sqrt{\frac{1776.86}{0.511}} = 58.97
\end{equation}
The square of the strange to down ratio has been multiplied by the ratio of largest to smallest eigenvalue in the electron family. There is about  0.6$\%$ deviation from the experimentally determined ratio.
\end{itemize}
It is interesting to note the pattern in which the eigenvalues multiply to give us root-mass ratios for various generations. The root-mass ratio of electron, up quark, and down quark is $\frac{1}{3}:\frac{2}{3}:1$. The root-mass ratio of
strange quark with respect to down quark is the ratio of the maximum eigenvalue to the minimum eigenvalue for charge 1. The ratio of bottom quark with respect to down quark can be similarly obtained but with an additional factor of maximum eigen value. Similar pattern is visible for up quark family and electron family, it is important to note that we use charge 1 eigenvalues for down quark family whereas charge $\frac{1}{3}$ eigenvalues are used for electron family. The root-mass ratios are summarised in the Fig. 1
\begin{figure}[h]
\centering
\includegraphics[width=16cm]{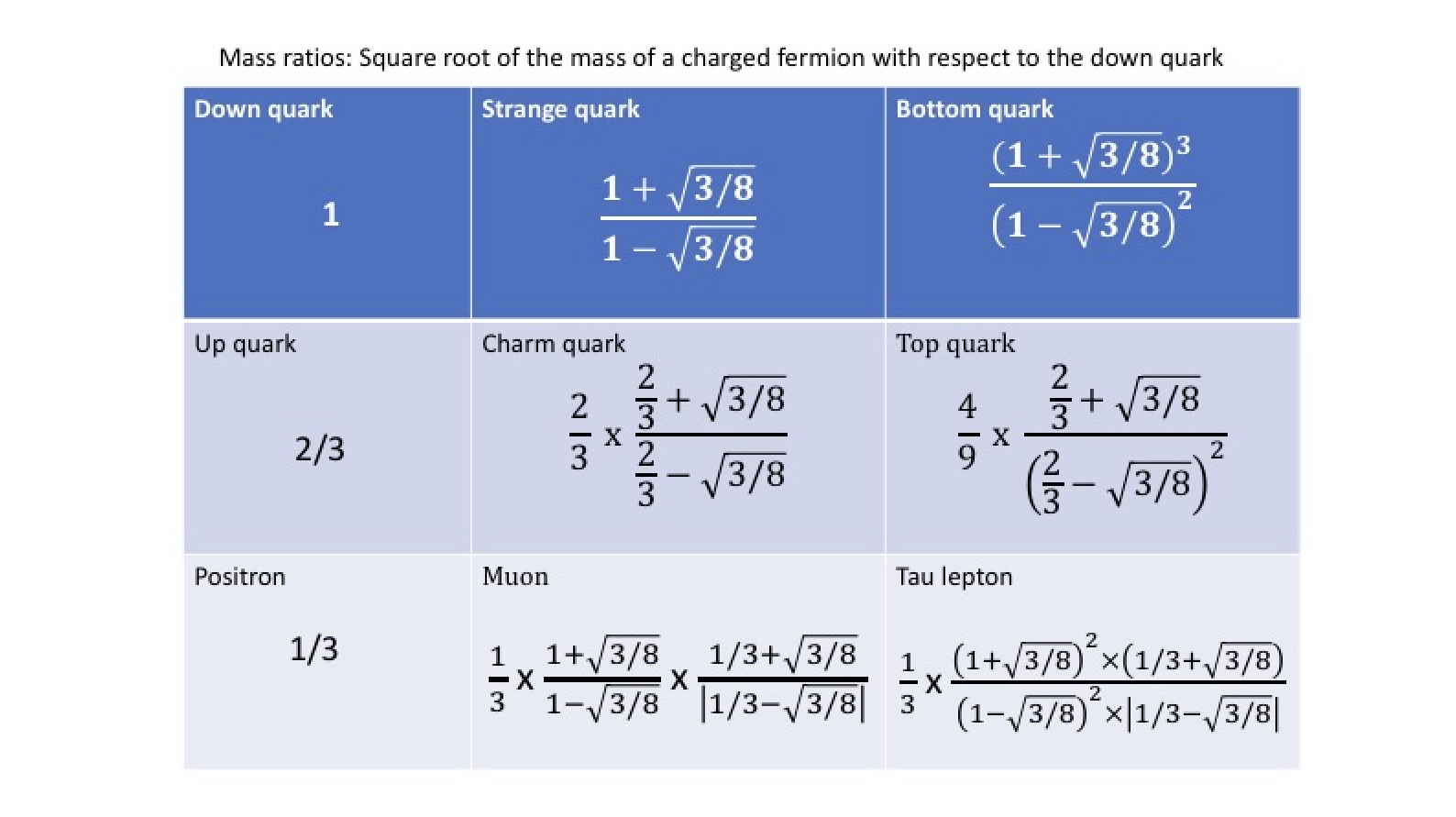}
\caption{Family-wise fermionic root-mass ratios assuming Majorana neutrino \cite{Tp3}}
\end{figure}
We are using charge 1 eigenvalues for the down quark family and charge $\frac{1}{3}$ eigenvalues for the electron family, this is because we are interpreting this charge as the mass number, more on this in section 8.2. An interesting thing to note is that the theoretically calculated root-mass ratios are lying within the experimental range considering error for the case of quarks, and depart $4\%$ or less for the charged leptons. Mass ratios for quarks are known more accurately from experiments  than their individual masses, and we will compare against such numbers in future work. Also, a deeper understanding as to why the square-root mass numbers are made in this specific way remains to be found.
\bigskip
\bigskip
\begin{center}
\begin{tabular}{ |p{3cm}||p{3cm}|p{3cm}|p{3cm}|  }
 \hline
 \multicolumn{4}{|c|}{Square root mass ratios} \\
 \hline
 Particles & Theoretical mass ratio & Minimum experimental value& Maximum experimental value\\
 \hline
 muon/electron   & 14.10    &  14.37913078& 14.37913090 \\
 taun/electron &   58.64  & 58.9660   &58.9700\\
 charm/up & 23.57& 21.04 &  26.87\\
 top/up &289.26 & 248.18&  310.07\\
 strange/down &   4.16  & 4.21 & 4.86\\
 bottom/down & 28.44  & 28.25   &30.97\\
 
 \hline
 \end{tabular}
\end{center}
\centerline{Table I: Comparison of theoretically predicted square-root mass ratio with experimentally known range}
\bigskip

Apart from the two mass ratios of charged leptons, other theoretical mass ratios lie within the experimental bounds \cite{Zyla1}. On accounting for the so-called Karolyhazy correction \cite{Tp3}  we might possibly get more accurate mass ratios for all particles including charged leptons.  This will be investigated in future work.

\subsection*{Root-mass ratios, assuming a Dirac neutrino, and using the corresponding Jordan eigenvalues}

If we had assumed the  neutrino to be a Dirac fermion, we would have obtained the following mass ratios, following the same construction as for the Majorana neutrino:
\begin{itemize}
\item Anti-strange quark with respect to anti-down quark
\begin{equation}
    \frac{1+\sqrt{3/2}}{1-\sqrt{3/2}} = 9.89; \quad \sqrt{\frac{95}{4.7}} = 4.50
\end{equation}
\item Anti-bottom quark with respect to anti-down quark
\begin{equation}
    \frac{1+\sqrt{3/2}}{1-\sqrt{3/2}}\times\frac{1+\sqrt{3/2}}{1}\times\frac{1+\sqrt{3/2}}{1-\sqrt{3/2}} = 218.00; \quad \sqrt{\frac{4180}{4.7}} = 29.82
\end{equation}
\item Charm quark with respect to up quark
\begin{equation}
    \frac{2/3+\sqrt{3/2}}{2/3-\sqrt{3/2}} = 3.39; \quad \sqrt{\frac{1275}{2.3}} = 23.55
\end{equation}
\item Top quark with respect to up quark
\begin{equation}
    \frac{2/3+\sqrt{3/2}}{2/3-\sqrt{3/2}}\times\frac{2/3}{2/3-\sqrt{3/2}} = 4.05; \quad \sqrt{\frac{173210}{2.3}} = 274.42
\end{equation}
\item Anti-muon with respect to electron
\begin{equation}
    \frac{1+\sqrt{3/2}}{1-\sqrt{3/2}}\times\frac{1/3+\sqrt{3/2}}{|1/3-\sqrt{3/2}|} = 17.30; \quad \sqrt{206.7682830} = 14.38
\end{equation}
\item Anti-tau lepton with respect to electron
\begin{equation}
    \frac{1+\sqrt{3/2}}{1-\sqrt{3/2}}\times\frac{1/3+\sqrt{3/2}}{|1/3-\sqrt{3/2}|}\times\frac{1+\sqrt{3/2}}{1-\sqrt{3/2}} = 171.27; \quad \sqrt{\frac{1776.86}{0.511}} = 58.97
\end{equation}
\end{itemize}
It can be seen clearly that the mass ratios are way off the experimental values in this case. In our previous work on the fine structure constant as well \cite{Tp3, GRFEssay2021}, it was essential to work with the Majorana neutrino, to obtain a theoretical value which agrees with experiment.   The derivation of  mass ratios strengthens our claim that the neutrino should be a Majorana particle and not a Dirac particle. A question may be asked on why a specific multiplication pattern for eigenvalues gives us the mass ratios. We do not know the answer to this question in entirety and further work is in progress in this regard.

\subsection{SU(3) Gravity}
It is interesting to note that the root mass ratio for positron, up quark, and anti-down quark is $\frac{1}{3}: \frac{2}{3}: 1$. Also it is clear that the ratio of charge of anti-down quark, up quark, and positron is exactly $\frac{1}{3}: \frac{2}{3}: 1$. In our work \cite{Vatsalya1}, we proposed a left-right symmetric model of fermions from $\mathbb{C}l(3)$ and $\mathbb{C}l(7)$,$\ \mathbb{C}l(7) = \mathbb{C}l(6) + \mathbb{C}l(6)$. In that paper we have  proposed that prior to L-R symmetry breaking we have left-handed fermions with $SU(3)$ color from one of the $\mathbb{C}l(6)$ whereas we have right-handed fermions with $SU(3)$ gravi-charge [i.e. root mass number]  from the other $\mathbb{C}l(6)$. For all left-handed electrically charged particles there is a right-handed particle with its fundamental  root mass number. For example the left-handed down quark has electric  charge $\frac{1}{3}$ along with a right-handed electron with a mass number $\frac{1}{3}$. These pair of particles will form a lepto-quark state prior to symmetry breaking. 

The exceptional group $F_4$ has two maximal subgroups $SU(3) \times SU(3)$ and $Spin(9)$, the intersection of these maximal subgroups is $SU(3)\times SU(2)\times U(1)$ which is the gauge group of the standard model. In our previous work \cite{Tp3}, we have shown that we can use the $SU(3)$ not lying in the intersection for getting three generations of fermions. Another interesting thing to note is that the complexified version of $F_4$ is the exceptional group $E_6$ which has two maximal subgroups $SU(3)\times SU(3)\times SU(3)$ and $Spin(10)$. The intersection of these two subgroups is $SU(3)\times SU(2)_L\times SU(2)_R\times U(1)$ which is that Pati-Salam gauge group for left-right symmetric model of fermions \cite{Boyle1}. Three generations of left-right symmetric fermions can be obtained from the $SU(3)$ not lying in the intersection of the two maximal subgroups. There is another $SU(3)$ not lying in the intersection, which we propose gives gravi-color to the lepto-quarks prior to L-R symmetry breaking. There will be three right-handed electrons with mass number $\frac{1}{3}$, the gravi-color acts only prior to symmetry breaking and in today's universe will be very weak because of the weak coupling constant of gravity.

In Section X below we discuss in some detail the prospects for unification of the standard model with gravity when the symmetry group of the theory is $E_6$, and the underlying space-time is eight dimensional octonionic space-time, not 4D Minkowski space-time.

\subsection{The Koide Formula}
The Koide formula for the experimentally measured masses of charged leptons is an unexplained empirical relation given by \cite{koide}
\begin{equation}
    \frac{m_e + m_{\mu} + m_{\tau}}{(\sqrt{m_e} + \sqrt{m_{\mu}} + \sqrt{m_{\tau}})^2} = 0.666661(7) \approx \frac{2}{3}
    \label{koide}
\end{equation}
We note that using our theoretical mass ratios we get the following theoretically predicted value
\begin{equation}
   \frac{m_e + m_{\mu} + m_{\tau}}{(\sqrt{m_e} + \sqrt{m_{\mu}} + \sqrt{m_{\tau}})^2}    = 0.669163 \approx \frac{2}{3}
\end{equation}
It remains to be seen if the Karolyhazy correction will predict an exact match between theory and experiment.

Another interesting point to note is that the eigen-values of charged leptons Jordan matrices for the Dirac neutrino case exactly satisfy the Koide formula
\begin{equation}
    \frac{(1+\sqrt{3/2})^2 + (1)^2 + (1-\sqrt(3/2))^2}{3^2} = \frac{2}{3}
    \label{koidedirac}
\end{equation}
This might be happening because prior to symmetry breaking the left-handed electrically charged fermions and right-handed fermions with mass charge come from Dirac neutrinos which post symmetery breaking become two distinct Majorana neutrinos with opposite chirality. 


Koide had also proposed  \cite{koide2} a relation for the Cabibbo angle $\theta_c$ in terms of masses of charged leptons:
\begin{equation}
\tan\theta_c = \sqrt{3} \frac{\sqrt{m_\mu} - \sqrt{m_e}}{2\sqrt{m_{\tau}}-\sqrt{m_\mu}-\sqrt{m_e}}=0.225
\end{equation} 
after using known mass values. Our theoretical mass ratios when used in the above formula give $\tan\theta_c=0.222$, whereas the experimental value for $\tan\theta_c$ is 0.22, thus there is very good agreement.  On the other hand,  if we assume the neutrino to be Dirac, we get $\tan\theta_c = 0.09$. This result encourages us to investigate the CKM mass matrix in connection with the Jordan eigenvalues - this will be taken up in future work.

\section{The Majorana Neutrino}

Recent experimental works \cite{ndbd} which discuss the possibility of the neutrinos being a Majorana particle give us further assurance that we are proceeding in the right direction, as by the calculations provided in the previous sections, the experimentally derived mass ratios clearly distinguish between the calculations done by once considering the neutrino to be a Dirac particle and again considering it to be a Majorana one. Interestingly, the same calculations which provide an excellent match for the Majorana case throw the values off by a large margin from those observed while  considering the neutrino to be a Dirac particle. It is entirely possible that a similar observation can be reproduced for a Dirac picture consisting of a different set of correlations between observed mass ratios to eigenvalue ratios and the reader is encouraged to explore it further. The authors of the present paper have tried to set up a consistent picture for the Dirac case like the one obtained for the Majorana one, but have been unable to draw any suggestive conclusion.

A further motivation for pursuing this claim is received from the paper \cite{Tp3} which discusses in detail as to how the eigenvalues obtained ultimately lead to the derivation of the value of the fine structure constant, which matches up to a very good accuracy, provided the neutrino is Majorana. Thus, it  adds to our confidence in the calculations, though we are still in the process of looking for a better set of relation rules for obtaining mass ratios which would be covariant with the quarks and leptons.

\section{Critique}
We discuss a few related aspects of our analysis  of mass ratios, which are currently under further development, and could provide additional insight into the results obtained above.

\subsection{Prospects for unification of the standard model with gravity when the symmetry group is $E_6$}
The discussion in this sub-section is motivated by the question: why is the square-root mass ratio 3: 2: 1 of the down quark, up quark and electron in the reverse order of their electric charge ratio  1: 2: 3? We believe that rather than being a coincidence, this fact points to deep physics, and that the symmetry group $E_6$ has an answer. We will assume that space-time is an eight dimensional manifold labeled by the octonions, and by virtue of the isomorphism $SL(2, \mathbb{O}) \sim SO(9,1)$ this is equivalent to a 10D Minkowski space-time manifold. Three generations of fermions reside on this space-time on which $E_6$ acting as the symmetry group is a candidate for the unification of the standard model with gravity, as we now argue. This in turn helps understand why the mass-ratios analysis works. The left-handed fermions are charge eigenstates, whereas the right-handed fermions are square-root mass eigenstates. The eigenvalues of the exceptional Jordan algebra, along with the corresponding eigen-matrices, permit the expression of charge eigenstates in terms of square-root mass eigenstates, and hence can be used to deduce mass ratios of charged fermions.

$E_6$ is the only exceptional Lie group which has complex representations, and it has two  maximal subgroups $\tilde{H_1} = [SU(3)\times SU(3)\times SU(3)]/\mathbf{Z_3}$, $\tilde{H_2} = Spin(10)$. Their intersection is $SU(3)\times SU(2)_R\times SU(2)_L\times U(1)$ which is the gauge-group for left-right symmetric model. The groups belonging to the two maximal sub-groups but lying outside the intersection are $Spin(6)$ and $SU(3)\times SU(3)$. We identify one of these two $SU(3)$ with generational symmetry, and now the novel part is that we introduce gravi-color, analogous to QCD color, and associate this third $SU(3)$ with gravitation and square-root mass number. This will help understand the down : up : electron square-root mass ratio of 3: 2: 1 Just as $SU(3)_c\times U(1)_{em}$ is described by the Clifford algebra $Cl(6)$ as unbroken electro-color, the group $SU(3)_{grav}\times U(1)_g$ will describe unbroken gravi-color through another copy of $Cl(6)$ and together these two copies of $Cl(6)$ will form a $Cl(7)$ using the complex split bioctonions \cite{Vatsalya1}. This offers a unification of QCD color with gravi-color, prior to the L-R symmetry breaking, which we assume is the same as the electro-weak symmetry breaking. The group $SU(2)_L \times SU(2)_R$ describes gravi-weak unification through complex split biquaternions; $SU(2)_L$ is the standard model weak symmetry and $SU(2)_R$ is the gravi- part of gravi-weak, mediated by two  gravitationally charged `Lorentz' bosons, a neutral Lorentz boson, and the Higgs. In our theory there are no right-handed weak bosons; these are replaced by three right-handed Lorentz bosons, and the electro-weak symmetry breaking also breaks the gravi-weak symmetry. The $Spin(6)$ which is not in the intersection is identified as a six dimensional Minkowski spacetime because of the isomorphism $Spin(6)\sim SO(5,1)\sim SL(2,H)$. This possibly is the space-time spanned by the gravi-weak interaction.

Prior to L-R symmetry breaking, the neutrino is a Dirac neutrino, which after symmetry breaking separates into the left-handed active Majorana neutrino, and the right-handed sterile Majorana neutrino. Analogous to how it was done in \cite{Vatsalya1}, we use the Dirac neutrino as an idempotent, prior to L-R symmetry breaking, and construct the Clifford algebra $Cl(7)=Cl(6) + Cl(6)$ displayed below.
\begin{align}
    {\mathcal{\overline V}_L} &= \frac{ie_8 + 1}{2} & \mathcal{\overline V}_R &= \frac{ie_8 + 1}{2}\\
    V_{ad1} &= \frac{(e_5 + ie_4)}{2} & V_{e+1} &= \omega\frac{(-e_5 - ie_4)}{2}\\
    V_{ad2} &= \frac{(e_3 + ie_1)}{2} & V_{e+2} &= \omega\frac{(-e_3 - ie_1)}{2}\\
    V_{ad3} &= \frac{(e_6 + ie_2)}{2} & V_{e+3} &= \omega\frac{(-e_6 - ie_2)}{2}\\
    V_{u1} &= \frac{(e_4 + ie_5)}{2} & V_{au1} &= \frac{(e_4 + ie_5)}{2}\\
    V_{u2} &= \frac{(e_1 + ie_3)}{2} & V_{au2} &= \frac{(e_1 + ie_3)}{2}\\
    V_{u3} &= \frac{(e_2 + ie_6)}{2} & V_{au3} &= \frac{(e_2 +ie_6)}{2}\\
    V_{e+} &= -\frac{(i + e_8)}{2} & V_{ad} &= \omega\frac{(i + e_8)}{2}
\end{align}
Notation is as in \cite{Vatsalya1}. The eight fermions on the left are made by using the left-handed anti-neutrino as the idempotent, while the eight fermions on the right are made by using the right-handed anti-neutrino as idempotent. The two sets share a common number  $U(1)_{electro-gravi}$ operator defined as usual by
\begin{equation}
     Q_{gem} = \frac{\alpha_1^{\dagger}\alpha_1 + \alpha_2^{\dagger}\alpha_2 + \alpha_3^{\dagger}\alpha_3}{3}
\end{equation}
and have an $SU(3)_c \times SU(3)_{grav}$ symmetry, which we interpret as the unification of QCD color and gravity, and also of electromagnetism and a $U(1)_{grav}$.  Here, $ Q_{gem}$  is the gravi-electric.charge number operator: after the symmetry breaking this will be interpreted as the electric charge for the left-handed particles, and square-root mass number for the right handed particles. The $U(1)_{electro-gravi}$ boson will separate into the photon for electromagnetism, and a newly proposed gravitational boson. Prior to symmetry breaking the particle content for one generation is as follows. Anti-particles are obtained by ordinary complex conjugation of the particles, as before.

The Dirac neutrino is the sum of the left handed neutrino and the right handed neutrino; it has $Q_{gem}=0$, is a singlet under $SU(3)_c \times SU(3)_{grav}$ and we can denote it as the particle LeftHandedNeutrino-RightHandedNeutrino, and after the L-R symmetry breaking it acquires mass and separates into a left-handed active Majorana neutrino and a right handed sterile Majorana neutrino.

The first excitation above the idempotent has $Q_{gem}=1/3$ and is an anti-triplet under $SU(3)_c$ and an anti-triplet under $SU(3)_{grav}$. We denote this particle as LeftHandedAntiDownQuark-RightHandedPositron. After the L-R symmetry breaking it separates into the left-handed anti-down quark of electric charge $1/3$ and right-handed positron of square-root mass number $1/3$ 

The second excitation above the idempotent has $Q_{gem}=2/3$ and is a triplet under $SU(3)_c$ and  a triplet under $SU(3)_{grav}$. We denote this particle as LeftHandedUpQuark-RightHandedUpQuark. After the L-R symmetry breaking it separates into the left-handed up quark of electric charge $2/3$ and right-handed up quark of square-root mass number $2/3$. 

The third excitation above the idempotent has $Q_{gem}=1$ and is a singlet under both $SU(3)_c$ and $SU(3)_{grav}$. We denote this particle as LeftHandedPositron-RightHandedAntiDownQuark. After the L-R symmetry breaking it separates into a left-handed positron of electric charge 1 and a right-handed anti-down quark of  square-root mass number 1.

The corresponding anti-particles have a $Q_{gem}$ number of the opposite sign.

We propose to identify the right-handed positron of square-root mass number 1/3 with the left-handed positron of electric charge 1 as being the same particle. This is essentially a proposal for a gauge-gravity duality which we hope to justify from the dynamics. Similarly, the right-handed anti-down quark with square-root mass number 1 is identified with the left-handed anti-down quark of electric charge 1/3. The right-handed up quark of square-root mass number 2/3 is identified with the left-handed up quark of electric charge 2/3. In this way we recover one generation of standard model fermions after the L-R symmetry breaking.

Before symmetry breaking, we can define $\ln \alpha_{unif}\propto 2\ln Q_{gem}\equiv \ln(AB) = \ln A + \ln B \propto e + \sqrt{m}$ where $\ln A$ is proportional to electric charge and $\ln B$ is proportional to square-root mass, and at the time of L-R symmetry breaking $2Q_{gem}$ separates into two equal parts, one identified with electric charge, and the other with square-root mass. We hence see that in the unified L-R phase we can define a new entity, a charge-root-mass as $\alpha_{unif} = \exp e \exp\sqrt{m}\equiv E\sqrt{M}$. This is the source of the unified force described by a $U(1)$ boson, sixteen gravi-gluons, and six gravi-weak bosons corresponding to $SU(2)_L \times SU(2)_R$ and the Higgs; adding to a total of 24 bosons. There are 48 fermions for three generations, giving a total of 48+24 =72, to which if we add six d.o.f. for the six dimensional space-time $SO(5,1)$ we might be able to account for the 78 dimensional $E_6$. The gravi-weak bosons generate the Lorentz-weak symmetry by their right action on the $Cl(7)$, as described in \cite{Vatsalya1}. After symmetry breaking this separates into the short range weak interaction and long-range gravity described by general relativity. $SU(3)_{grav}$ is negligible in strength  compared to QCD color but plays a very important role of describing the square-root mass number as source of would-be-gravity and showing that mass-quantisation arises only after the standard model has been unified with gravity, as was always anticipated. We also see via $E_6$ that $SU(3)_{grav} \times SU(2)_R \times U(1)_g$ is the gravitational counterpart of the standard model $SU(3)_{c} \times SU(2)_L \times U(1)_{em}$. The remaining entities from the two maximal sub-groups, i.e. $SU(3)_{gen}$ and $Spin(6)$ respectively give rise to three generations and a 6D Minkowski space-time.
We now finally understand why the square-root mass ratios 3:2:1 for down : up : electron are in the reverse order as the ratio $1 : 2 : 3$ of their electric charge. It is a consequence of the gauge-gravity duality afforded by $E_6$.

\subsection{Outlook}

A careful look at the analysis we have presented in this paper could raise further questions, including aspects which yet remain to be resolved. Below we discuss a few such issues in a systematic manner:

\begin{itemize}

\item We noted the fact that the observed square-root mass ratio of positron, up quark
and anti-down quark is nearly 1:2:3, which is in the reverse order of their electric charge
ratio 3:2:1. This coincidence motivated us to relate  gravity to the Standard Model and
establish the gauge-gravity duality under a larger symmetry group $E_6$. We assumed a left-
right symmetry and a common number $U(1)$ operator $Q_{gem}$ interpreted as the gravi-electric
charge number operator. After the left-right symmetry breaking, $Q_{gem}$ will be interpreted as
the electric-charge for left-handed particles and square-root mass number for right-handed
particles. This seems to explain the inverse relation between the square-root mass ratio and
the electric charge ratio. However, the electric charge and the gravity charge (i.e., the mass)
exhibit very differently in physics. The most signicant distinction is that the electric charge
of a particle is protected by the $U(1)_{em}$ gauge symmetry and thus free from the radiative
corrections. In contrast, the mass of a particle (especially for light quarks, which do not
have a well-defined pole mass) may run with the energy scale. In other words, the 1:2:3
square-root mass ratio of positron, up quark and anti-down quark will be violated by the
radiative corrections in general. Therefore, the naive unication of the electric charge and
gravity charge by a common $U(1)$ charge before the symmetry breaking might be incorrect?

From a quantum field theoretic point of view,  radiative corrections will indeed disrupt the square-root mass relation 1:2:3 However, this relation is not intended or implied to be true at all energy scales. Furthermore, the question of validity of this relation must be decoupled from energy scale. This particular square-root-mass relation is true when the electron can be treated as reaching the no-interaction limit [this happens at low energies] and the down quark and up quark can be treated as reaching the no-interaction limit [this happens at high energies]. Thus the relation 1:2:3 for square-root mass ratios is defined for when the electron is at low energies, and the down and up quark are at the high energy asymptotic freedom limit. Any departure from this limit, either for the electron, or for the quarks, will cause a deviation from the ratio 1:2:3 However such a deviation is consistent with and as expected from quantum field theory, and not a problem for the octonionic theory. We have calculated the mass ratios for the situation when the corresponding particles reach their interaction free limit, and the fact that for this to happen more than one energy scale is involved is not a problem.

\item Although the assumption of Majorana nature of neutrinos can reproduce the correct mass
ratios for charged-fermions, it cannot accommodate the tiny but nonzero masses of neutrinos,
which have been  firmly proved by the neutrino oscillation experiments. Worse still, the
gauge-gravity duality established in the octonionic theory does not hold for neutrinos because
of their electric neutrality but nonzero masses. So how to explain neutrino masses in the
framework of octonions and exceptional Jordan algebra?

This is a point of great importance, and an acid test for the octonionic theory as to whether eventually it can predict neutrino masses and mass ratios. This is a task for the future and work is in progress in this direction. However we can make the following important observation: even for the neutrino, which has zero electric charge, all the Jordan eigenvalues are not zero. In fact for the case of the Dirac neutrino, relevant before L-R symmetry, none of the three eigenvalues are zero, these being $(-1/2-\sqrt{3}/2, 1, -1/2 + \sqrt{3}/2)$. For the Majorana neutrino case, relevant after L-R symmetry breaking, only one of the three eigenvalues is zero, the other two eigenvalues being $(\sqrt{3}/2, -\sqrt{3}/2)$. The fact that even for zero electric charge there are non-zero eigenvalues indicates that in this theory neutrinos will have mass, though the mechanism of acquiring mass remains to be understood. Subject to further analysis we can speculate that the three right-handed sterile neutrinos will have the same mass as their corresponding same generation left-handed active neutrino counterpart. And that two out of the six neutrinos are massless, four have mass. We note the fundamental difference between charged fermions and the neutrinos: the former all experience both the weak force as well as gravity; whereas the active neutrino does experience both the forces, but the sterile neutrino only gravity.

\item It is miraculous that the eigenvalues of exceptional Jordan algebra can reproduce the almost
correct mass ratios of charged-fermions in the Standard Model. Is it just a coincidence or
there is any profound connection between the mathematics and physics therein?

As we saw in the previous section, the ability of the exceptional Jordan algebra to explain mass ratios of charged fermions arises from  a strong physical motivation. Namely that elementary particles should fundamentally be described as living in a non-commutative spinor spacetime, not in a 4D Minkowski spacetime, this latter only being an approximate description. The sought for exact description in a spinor spacetime is achieved in octonion space, and by extending the standard model to include a right-handed sector which describes `would-be-gravity'. When this is done, quantisation of electric charge and square-root mass is an inevitable outcome. This can also be called  a relativistic weak quantum gravity effect on the standard model, and we realise that unification of gravity and the standard model is essential at all energy scales, not just at the Planck energy scale. There is an associated dynamics, known as generalised trace dynamics, from which quantum theory and gravitation are both emergent. The fact that mass ratios are derived nearly correctly, alongside the fine structure constant, are likely indicators that this theory is on the right track \cite{Singhreview}. 

Furthermore, right-handed sterile neutrinos arise unavoidably, in the extension to include the right-handed gravitational sector. Sterile neutrinos interact with other particles only via the gravitational force. Hence, as soon we include them in the standard model, we bring in gravity. And since the standard model can only be described and understood in a quantum setting, by bringing in sterile neutrinos we bring in quantum gravity, and unification. Hence, any extension of the standard model which includes sterile neutrinos must also present a consistent theory of quantum gravity and unification. Only after that has been achieved, can theorists present experimentalists with unambiguous sterile neutrino signatures to look for. In that sense too, the octonionic theory holds out promise, and it's implications for neutrino experiments should be studied carefully.

\item The 8-dimensional octonionic manifold is equivalent to the 10-dimensional Minkowski space-
time due to the mathematical fact that $SL (2;\mathbb O)$ is the double cover of $SO(9; 1)$. This space-
time dimension happens to be the one predicted by string theory. So is there any relationship
between octonions and string theory?

Indeed there is, and perhaps it is reasonable to suggest that the octonionic theory is an improvement over string theory which resolves the difficulties of the latter, transforming it into a predictable and falsifiable theory. By demanding that there exist a reformulation of quantum field theory which does not depend on classical time, we arrive at a pre-spacetime pre-quantum matrix-valued Lagrangian dynamics of two dimensional extended objects. These entities, which we call `atoms' of spacetime-matter or aikyons, are strongly reminiscent of the strings of string theory, as all elementary particles are excitations of the aikyon. The principal differences from string theory are the following. Elementary particles are defined on the spinorial octonionic space - equivalent to 10D Minkowski spacetime -  evolving in the absolute Connes time. This immediately reveals the standard model. Furthermore, this Lagrangian dynamics is not quantised, but is already pre-quantum. From here, quantum field theory and gravitation are emergent. Also, the Hamiltonian of the theory is not self-adjoint in general. If the fermions in the theory achieve a critical degree of entanglement, the anti-self-adjoint part of the Hamiltonian becomes significant, resulting in spontaneous localisation and the emergence of 4D classical spacetime and macroscopic classical objects which are confined to four spacetime dimensions. This is compactification without compactification. Because those systems which have not achieved critical entanglement - for them the anti-self-adjoint part of the Hamiltonian is negligible and they obey the emergent laws of quantum theory - continue to live in ten spacetime dimensions. The extra dimensions are never compactified in an ad hoc manner, unlike in string theory (where ad hoc compactification leads to the serious problems of non-uniqueness, non-falsifiability and non-predictability). The thickness of these extra dimensions is not Planck length, but is rather determined by the support of the wave function of the system under consideration. 

We believe that the octonionic theory is a way of arriving at  a refined and now successful formulation of string theory, by starting from foundational motivations. We do not start by proposing that elementary particles are described by extended objects i.e. strings, and that the quantum theory of strings is a theory of unification. 

\item Can the Jordan eigenvalues reproduce the correct 
flavor mixing angles and CP-violating
phases in the quark and leptonic sector?

This is currently work in progress. We are investigating if the twelve horizontal Jordan eigenvalues between themselves determine the 25 dimensionless constants of the standard model. 

\end{itemize}

\section{Conclusions}
We would like to conclude that we can obtain three generations of fermions by rotating the first generation in the octonionic space. This rotation is due to the unaccounted $SU(3)$ symmetry group present in the $F_4$ group. On writing the three generations of fermions in a $3\times 3$ matrix with diagonal entries for electric charge we obtain exceptional Jordan matrices and we calculate its eigenvalues. The eigenvalues remain same even if we choose some other color for the quarks, or even if we work with anti-particles in place of particles. We conclude that these eigenvalues are simultaneously related to the electric charge and mass for a type of particle across the generations. Using these eigenvalues we calculate the mass ratios of fermions for anti-down quark, up quark, and electron family. We show that these mass ratios hold true if we consider the neutrino to be Majorana instead of Dirac. Our previous work on the calculation of fine structure constant also suggests the neutrino to be a Majorana fermion. We have also shown the eigenmatrices in this paper along with the eigenvalues. These eigenmatrices can play an important role in understanding the three generations problem. We also discuss root-mass numbers as a fundamental quantum number analogous to the electric charge. This root-mass number comes from another unaccounted $SU(3)$ group in the $E_6$ group, and this $SU(3)$ gives us gravi-color which is very weak because of the weak coupling constant of gravity. $SU(3)$ gravity also explains the root-mass ratio of $\frac{1}{3}, \frac{2}{3}, 1$ for the electron, up quark, and down quark. 

\section{Appendix: Quaternionic eigenmatrices corresponding to the Jordan eigenvalues}
The Jordan Eigenvalue Problem has been dealt with extensively in previous literature. Dray and Manogue, for instance, utilized the Jordan product $\mathcal{A}\circ \mathcal{B} = \frac{1}{2}(\mathcal{A}\mathcal{B}+\mathcal{B}\mathcal{A})$ to obtain the eigenmatrices \cite{Dray1} corresponding to calculated eigenvalues. They observed that an octonionic matrix $\mathcal{A}$ can be written so as to decompose into its eigenmatrices $\mathcal{P}_{\lambda}$ as
\begin{equation}
    \mathcal{A} = \sum_{i=1}^3 \lambda_i \mathcal{P}_{\lambda_i}
\end{equation}
Even though $\mathcal{A}$ is a matrix with octonionic entries, the $\mathcal{P}_{\lambda_i}$ lie in quaternionic subalgebras, which we have demonstrated below. The exact physical interpretation of these eigenmatrices in terms of the mass eigenstates for individual particles is under further investigation.

For the Jordan matrix $$X=
\begin{bmatrix}
    q & a & \Tilde{b} \\
    \Tilde{a} & q & c\\
    b & \Tilde{c} & q
\end{bmatrix}$$
we get the eigenmatrix of the form
\begin{equation}
    \mathcal{P}_{\lambda}=\frac{1}{3\lambda^{'2}-(a\Tilde{a}+b\Tilde{b}+c\Tilde{c})}
    \begin{bmatrix}
        \lambda^{'2}-c\Tilde{c}\; & \Tilde{b}\Tilde{c}-\lambda'a\; & ac-\lambda'\Tilde{b}\\
        cb-\lambda'\Tilde{a} & \lambda^{'2}-b\Tilde{b} & \Tilde{a}\Tilde{b}-\lambda'c\\
        \Tilde{c}\Tilde{a}-\lambda'b & ba-\lambda'\Tilde{c} & \lambda^{'2}-a\Tilde{a}
    \end{bmatrix}  
\end{equation}
where $\lambda'=q-\lambda$

\noindent {\it Majorana Neutrino Set:}
For the case of the Majorana neutrino, we reduce the eigenmatrices $\mathcal{P}_{\lambda_i}$ to their octonionic coordinates, to show that each eigenmatrix lies in the quaternionic subalgebra determined by its original family.

\textbf{Neutrino \((V_{\nu)}\)}
\begin{equation}
 \mathcal{P}_0 =
    \begin{bmatrix} 
        \frac{1}{3} & \frac{1}{3} & \frac{-\sqrt{3}e_6-1}{6} \\ \frac{1}{3} & \frac{1}{3} & \frac{-\sqrt{3}e_6-1}{6} \\ \frac{\sqrt{3}e_6-1}{6} & \frac{\sqrt{3}e_6-1}{6} & \frac{1}{3}
    \end{bmatrix}\qquad
    \mathcal{P}_{\frac{\sqrt{3}}{2}} =
    \begin{bmatrix} 
        \frac{1}{3} & \frac{\sqrt{3}e_6-1}{6} & \frac{\sqrt{3}e_6-1}{6} \\ 
        \frac{-\sqrt{3}e_6-1}{6} & \frac{1}{3} & \frac{-\sqrt{3}e_6+1}{6} \\ 
        \frac{-\sqrt{3}e_6-1}{6} & \frac{\sqrt{3}e_6+1}{6} & \frac{1}{3}
    \end{bmatrix}\qquad
    \mathcal{P}_{-\frac{\sqrt{3}}{2}} =
    \begin{bmatrix} 
        \frac{1}{3} & \frac{-\sqrt{3}e_6-1}{6} & \frac{1}{3} \\ 
        \frac{\sqrt{3}e_6-1}{6} & \frac{1}{3} & -\frac{1}{3} \\
        \frac{1}{3} & -\frac{1}{3} & \frac{1}{3}
    \end{bmatrix}
    \end{equation}

\textbf{Anti-down quark \((V_{ad})\)}
\begin{equation}
    \mathcal{P}_{1} =
    \begin{bmatrix}
        \frac{1}{3} & \frac{1-\sqrt{3}e_3+\sqrt{3}e_5+3e_2}{12} & \frac{-1-\sqrt{3}e_5}{6} \\ \frac{1+\sqrt{3}e_3-\sqrt{3}e_5-3e_2}{12} & \frac{1}{3} & \frac{-1-\sqrt{3}e_3}{6} \\ 
        \frac{-1+\sqrt{3}e_5}{6} & \frac{-1+\sqrt{3}e_3}{6} & \frac{1}{3}
    \end{bmatrix}\notag
\end{equation}
\begin{equation}
    \mathcal{P}_{1+\sqrt{\frac{3}{8}}} =
    \begin{bmatrix}
        \frac{1}{3} & \frac{-1+(1+\sqrt{2})\sqrt{3}e_3+(-1+\sqrt{2})\sqrt{3}e_5-3e_2}{24} & \frac{(\sqrt{2}-\sqrt{3})+e_3+(\sqrt{6}+1)e_5-\sqrt{3}e_2}{12\sqrt{2}} \\
        \frac{-1+(-1-\sqrt{2})\sqrt{3}e_3+(1-\sqrt{2})\sqrt{3}e_5+3e_2}{24} & \frac{1}{3} & \frac{(\sqrt{2}+\sqrt{3})+(\sqrt{6}-1)e_3-e_5-\sqrt{3}e_2}{12\sqrt{2}} \\
        \frac{(\sqrt{2}-\sqrt{3})-e_3+(-\sqrt{6}-1)e_5+\sqrt{3}e_2}{12\sqrt{2}} & \frac{(\sqrt{2}+\sqrt{3})+(-\sqrt{6}+1)e_3+e_5+\sqrt{3}e_2}{12\sqrt{2}} & \frac{1}{3}
    \end{bmatrix}
\end{equation}
\begin{equation}
    \mathcal{P}_{1-\sqrt{\frac{3}{8}}} =
    \begin{bmatrix}
        \frac{1}{3} & \frac{-1+(1-\sqrt{2})\sqrt{3}e_3+(-1-\sqrt{2})\sqrt{3}e_5-3e_2}{24} & \frac{(\sqrt{2}+\sqrt{3})-e_3+(\sqrt{6}-1)e_5+\sqrt{3}e_2}{12\sqrt{2}} \\
        \frac{-1+(-1+\sqrt{2})\sqrt{3}e_3+(1+\sqrt{2})\sqrt{3}e_5+3e_2}{24} & \frac{1}{3} & \frac{(\sqrt{2}-\sqrt{3})+(\sqrt{6}+1)e_3+e_5+\sqrt{3}e_2}{12\sqrt{2}} \\
        \frac{(\sqrt{2}+\sqrt{3})+e_3+(-\sqrt{6}+1)e_5-\sqrt{3}e_2}{12\sqrt{2}} & \frac{(\sqrt{2}-\sqrt{3})+(-\sqrt{6}-1)e_3-e_5-\sqrt{3}e_2}{12\sqrt{2}} & \frac{1}{3}
    \end{bmatrix}\notag
\end{equation}

\textbf{Up quark \((V_{u})\)}
\begin{equation}
    \mathcal{P}_{\frac{2}{3}} =
    \begin{bmatrix}
        \frac{1}{3} & \frac{1-\sqrt{3}e_4+\sqrt{3}e_2-3e_1}{12} & \frac{-1-\sqrt{3}e_2}{6} \\ \frac{1+\sqrt{3}e_4-\sqrt{3}e_2+3e_1}{12} & \frac{1}{3} & \frac{-1-\sqrt{3}e_4}{6} \\ 
        \frac{-1+\sqrt{3}e_2}{6} & \frac{-1+\sqrt{3}e_4}{6} & \frac{1}{3}
    \end{bmatrix}\notag
\end{equation}
\begin{equation}
    \mathcal{P}_{\frac{2}{3}+\sqrt{\frac{3}{8}}} =
    \begin{bmatrix}
        \frac{1}{3} & \frac{-1+(1+\sqrt{2})\sqrt{3}e_4+(-1+\sqrt{2})\sqrt{3}e_2+3e_1}{24} & \frac{(\sqrt{2}-\sqrt{3})+e_4+(\sqrt{6}+1)e_2+\sqrt{3}e_1}{12\sqrt{2}} \\
        \frac{-1+(-1-\sqrt{2})\sqrt{3}e_4+(1-\sqrt{2})\sqrt{3}e_2-3e_1}{24} & \frac{1}{3} & \frac{(\sqrt{2}+\sqrt{3})+(\sqrt{6}-1)e_4-e_2+\sqrt{3}e_1}{12\sqrt{2}} \\
        \frac{(\sqrt{2}-\sqrt{3})-e_4+(-\sqrt{6}-1)e_2-\sqrt{3}e_1}{12\sqrt{2}} & \frac{(\sqrt{2}+\sqrt{3})+(-\sqrt{6}+1)e_4+e_2-\sqrt{3}e_1}{12\sqrt{2}} & \frac{1}{3}
    \end{bmatrix}
\end{equation}
\begin{equation}
    \mathcal{P}_{\frac{2}{3}-\sqrt{\frac{3}{8}}} =
    \begin{bmatrix}
        \frac{1}{3} & \frac{-1+(1-\sqrt{2})\sqrt{3}e_4+(-1-\sqrt{2})\sqrt{3}e_2+3e_1}{24} & \frac{(\sqrt{2}+\sqrt{3})-e_4+(\sqrt{6}-1)e_2-\sqrt{3}e_1}{12\sqrt{2}} \\
        \frac{-1+(-1+\sqrt{2})\sqrt{3}e_4+(1+\sqrt{2})\sqrt{3}e_2-3e_1}{24} & \frac{1}{3} & \frac{(\sqrt{2}-\sqrt{3})+(\sqrt{6}+1)e_4+e_2-\sqrt{3}e_1}{12\sqrt{2}} \\
        \frac{(\sqrt{2}+\sqrt{3})+e_4+(-\sqrt{6}+1)e_2+\sqrt{3}e_1}{12\sqrt{2}} & \frac{(\sqrt{2}-\sqrt{3})+(-\sqrt{6}-1)e_4-e_2+\sqrt{3}e_1}{12\sqrt{2}} & \frac{1}{3}
    \end{bmatrix}\notag
\end{equation}
\textbf{Positron \((V_{e^{+}})\)}
\begin{equation}
    \mathcal{P}_{\frac{1}{3}} =
    \begin{bmatrix}
        \frac{1}{3} & \frac{1+\sqrt{3}e_1-\sqrt{3}e_7+3e_3}{12} & \frac{-1-\sqrt{3}e_7}{6} \\ \frac{1-\sqrt{3}e_1+\sqrt{3}e_7-3e_3}{12} & \frac{1}{3} & \frac{-1-\sqrt{3}e_1}{6} \\ 
        \frac{-1+\sqrt{3}e_7}{6} & \frac{-1+\sqrt{3}e_1}{6} & \frac{1}{3}
    \end{bmatrix}\notag
\end{equation}
\begin{equation}
    \mathcal{P}_{\frac{1}{3}+\sqrt{\frac{3}{8}}} =
    \begin{bmatrix}
        \frac{1}{3} & \frac{-1+(-1+\sqrt{2})\sqrt{3}e_1+(1+\sqrt{2})\sqrt{3}e_7-3e_3}{24} & \frac{(\sqrt{2}-\sqrt{3})+e_1+(\sqrt{6}+1)e_7-\sqrt{3}e_3}{12\sqrt{2}} \\
        \frac{-1+(1-\sqrt{2})\sqrt{3}e_1+(-1-\sqrt{2})\sqrt{3}e_7+3e_3}{24} & \frac{1}{3} & \frac{(\sqrt{2}+\sqrt{3})+(\sqrt{6}-1)e_1-e_7-\sqrt{3}e_3}{12\sqrt{2}} \\
        \frac{(\sqrt{2}-\sqrt{3})-e_1+(-\sqrt{6}-1)e_7+\sqrt{3}e_3}{12\sqrt{2}} & \frac{(\sqrt{2}+\sqrt{3})+(-\sqrt{6}+1)e_1+e_7+\sqrt{3}e_3}{12\sqrt{2}} & \frac{1}{3}
    \end{bmatrix}
\end{equation}
\begin{equation}
    \mathcal{P}_{\frac{1}{3}-\sqrt{\frac{3}{8}}} =
    \begin{bmatrix}
        \frac{1}{3} & \frac{-1+(-1-\sqrt{2})\sqrt{3}e_1+(1-\sqrt{2})\sqrt{3}e_7-3e_3}{24} & \frac{(\sqrt{2}+\sqrt{3})-e_1+(\sqrt{6}-1)e_7+\sqrt{3}e_3}{12\sqrt{2}} \\
        \frac{-1+(1+\sqrt{2})\sqrt{3}e_1+(-1+\sqrt{2})\sqrt{3}e_7+3e_3}{24} & \frac{1}{3} & \frac{(\sqrt{2}-\sqrt{3})+(\sqrt{6}+1)e_1+e_7+\sqrt{3}e_3}{12\sqrt{2}} \\
        \frac{(\sqrt{2}+\sqrt{3})+e_1+(-\sqrt{6}+1)e_7-\sqrt{3}e_3}{12\sqrt{2}} & \frac{(\sqrt{2}-\sqrt{3})+(-\sqrt{6}-1)e_1-e_7-\sqrt{3}e_3}{12\sqrt{2}} & \frac{1}{3}
    \end{bmatrix}\notag
\end{equation}
Here, we make two rather interesting observations. First, all the diagonal entries are $\frac{1}{3}$, which corresponds to the lowest quantized charge of the antidown quark. Secondly, due to its original octonionic representations, the neutrino is once again limited to only one imaginary basis along with unity, as opposed to the charged fermions which are characterised by a unique quaternionic subalgebra. We comment on the latter further in this paper.

\noindent{\it Dirac Neutrino Set:}
Along parallel lines, we find the eigenmatrices given the assumption that the neutrino is a Dirac particle and get the following results

\textbf{Neutrino ($V_{\nu}$)}
\begin{equation}
    \mathcal{P}_1 = 
    \begin{bmatrix}
        \frac{1}{3} & \frac{2}{3}(\Tilde{V}_{\mu\nu}\Tilde{V}_{\tau\nu}+V_{\nu}) & \frac{2}{3}(V_{\nu}V_{\tau\nu}+\Tilde{V}_{\mu\nu})\\
        \frac{2}{3}(V_{\tau\nu}V_{\mu\nu}+\Tilde{V}_{\nu}) & \frac{1}{3} & \frac{2}{3}(\Tilde{V}_{\nu}\Tilde{V}_{\mu\nu}+V_{\tau\nu})\\
        \frac{2}{3}(\Tilde{V}_{\tau\nu}\Tilde{V}_{\nu}+V_{\mu\nu}) & \frac{2}{3}(V_{\mu\nu}V_{\nu}+\Tilde{V}_{\tau\nu}) & \frac{1}{3}
    \end{bmatrix}\notag
\end{equation}
\begin{equation}
    \mathcal{P}_{\frac{-1-\sqrt{3}}{2}} = 
    \begin{bmatrix}
        \frac{1}{3} & \frac{1}{3}(\frac{2\Tilde{V}_{\mu\nu}\Tilde{V}_{\tau\nu}}{-1-\sqrt{3}}+V_{\nu}) & \frac{1}{3}(\frac{2V_{\nu}V_{\tau\nu}}{-1-\sqrt{3}}+\Tilde{V}_{\mu\nu})\\
        \frac{1}{3}(\frac{2V_{\tau\nu}V_{\mu\nu}}{-1-\sqrt{3}}+\Tilde{V}_{\nu}) & \frac{1}{3} & \frac{1}{3}(\frac{2\Tilde{V}_{\nu}\Tilde{V}_{\mu\nu}}{-1-\sqrt{3}}+V_{\tau\nu})\\
        \frac{1}{3}(\frac{2\Tilde{V}_{\tau\nu}\Tilde{V}_{\nu}}{-1-\sqrt{3}}+V_{\mu\nu}) & \frac{1}{3}(\frac{2V_{\mu\nu}V_{\nu}}{-1-\sqrt{3}}+\Tilde{V}_{\tau\nu}) & \frac{1}{3}
    \end{bmatrix}
\end{equation}
\begin{equation}
    \mathcal{P}_{\frac{1+\sqrt{3}}{2}} = 
    \begin{bmatrix}
        \frac{1}{3} & \frac{1}{3}(\frac{2\Tilde{V}_{\mu\nu}\Tilde{V}_{\tau\nu}}{1+\sqrt{3}}+V_{\nu}) & \frac{1}{3}(\frac{2V_{\nu}V_{\tau\nu}}{1+\sqrt{3}}+\Tilde{V}_{\mu\nu})\\
        \frac{1}{3}(\frac{2V_{\tau\nu}V_{\mu\nu}}{1+\sqrt{3}}+\Tilde{V}_{\nu}) & \frac{1}{3} & \frac{1}{3}(\frac{2\Tilde{V}_{\nu}\Tilde{V}_{\mu\nu}}{1+\sqrt{3}}+V_{\tau\nu})\\
        \frac{1}{3}(\frac{2\Tilde{V}_{\tau\nu}\Tilde{V}_{\nu}}{1+\sqrt{3}}+V_{\mu\nu}) & \frac{1}{3}(\frac{2V_{\mu\nu}V_{\nu}}{1+\sqrt{3}}+\Tilde{V}_{\tau\nu}) & \frac{1}{3}
    \end{bmatrix}\notag
\end{equation}
\noindent \textbf{Antidown quark ($V_{ad}$)}
\begin{equation}
    \mathcal{P}_1 = \begin{bmatrix}
        \frac{1}{3} & -\frac{2}{3}\Tilde{V}_{as}\Tilde{V}_{ab} & -\frac{2}{3}V_{ad}V_{ab}\\
        -\frac{2}{3}V_{ab}V_{as} & \frac{1}{3} & -\frac{2}{3}\Tilde{V}_{ad}\Tilde{V}_{as}\\
        -\frac{2}{3}\Tilde{V}_{ab}\Tilde{V}_{ad} & -\frac{2}{3}V_{as}V_{ad} & \frac{1}{3}
    \end{bmatrix}\notag
\end{equation}
\begin{equation}
    \mathcal{P}_{1\pm \sqrt{\frac{3}{2}}} = \begin{bmatrix}
        \frac{1}{3} & \frac{1}{3}(\Tilde{V}_{as}\Tilde{V}_{ab}\pm \sqrt{\frac{3}{2}}V_{ad}) & \frac{1}{3}(V_{ad}V_{ab}\pm \sqrt{\frac{3}{2}}\Tilde{V}_{as})\\
        \frac{1}{3}(V_{ab}V_{as}\pm \sqrt{\frac{3}{2}}\Tilde{V}_{ad}) & \frac{1}{3} & \frac{1}{3}(\Tilde{V}_{ad}\Tilde{V}_{as}\pm \sqrt{\frac{3}{2}}V_{ab})\\
        \frac{1}{3}(\Tilde{V}_{ab}\Tilde{V}_{ad}\pm \sqrt{\frac{3}{2}}V_{as}) & \frac{1}{3}(V_{as}V_{ad}\pm \sqrt{\frac{3}{2}}\Tilde{V}_{ab}) & \frac{1}{3}
    \end{bmatrix}
\end{equation}
\noindent \textbf{Up quark ($V_u$)}
\begin{equation}
    \mathcal{P}_{\frac{2}{3}} = \begin{bmatrix}
        \frac{1}{3} & -\frac{2}{3}\Tilde{V}_{c}\Tilde{V}_{t} & -\frac{2}{3}V_{u}V_{t}\\
        -\frac{2}{3}V_{t}V_{c} & \frac{1}{3} & -\frac{2}{3}\Tilde{V}_{u}\Tilde{V}_{c}\\
        -\frac{2}{3}\Tilde{V}_{t}\Tilde{V}_{u} & -\frac{2}{3}V_{c}V_{u} & \frac{1}{3}
    \end{bmatrix}\notag
\end{equation}
\begin{equation}
    \mathcal{P}_{\frac{2}{3}\pm \sqrt{\frac{3}{2}}} = \begin{bmatrix}
        \frac{1}{3} & \frac{1}{3}(\Tilde{V}_{c}\Tilde{V}_{t}\pm \sqrt{\frac{3}{2}}V_{u}) & \frac{1}{3}(V_{u}V_{t}\pm \sqrt{\frac{3}{2}}\Tilde{V}_{c})\\
        \frac{1}{3}(V_{t}V_{c}\pm \sqrt{\frac{3}{2}}\Tilde{V}_{u}) & \frac{1}{3} & \frac{1}{3}(\Tilde{V}_{u}\Tilde{V}_{c}\pm \sqrt{\frac{3}{2}}V_{t})\\
        \frac{1}{3}(\Tilde{V}_{t}\Tilde{V}_{u}\pm \sqrt{\frac{3}{2}}V_{c}) & \frac{1}{3}(V_{c}V_{u}\pm \sqrt{\frac{3}{2}}\Tilde{V}_{t}) & \frac{1}{3}
    \end{bmatrix}
\end{equation}
\noindent \textbf{Positron ($V_{e^+}$)}
\begin{equation}
    \mathcal{P}_{\frac{1}{3}} = \begin{bmatrix}
        \frac{1}{3} & -\frac{2}{3}\Tilde{V}_{a\mu}\Tilde{V}_{a\tau} & -\frac{2}{3}V_{e^+}V_{a\tau}\\
        -\frac{2}{3}V_{a\tau}V_{a\mu} & \frac{1}{3} & -\frac{2}{3}\Tilde{V}_{e^+}\Tilde{V}_{a\mu}\\
        -\frac{2}{3}\Tilde{V}_{a\tau}\Tilde{V}_{e^+} & -\frac{2}{3}V_{a\mu}V_{e^+} & \frac{1}{3}
    \end{bmatrix}\notag
\end{equation}
\begin{equation}
    \mathcal{P}_{\frac{1}{3}\pm \sqrt{\frac{3}{2}}} = \begin{bmatrix}
        \frac{1}{3} & \frac{1}{3}(\Tilde{V}_{a\mu}\Tilde{V}_{a\tau}\pm \sqrt{\frac{3}{2}}V_{e^+}) & \frac{1}{3}(V_{e^+}V_{a\tau}\pm \sqrt{\frac{3}{2}}\Tilde{V}_{a\mu})\\
        \frac{1}{3}(V_{a\tau}V_{a\mu}\pm \sqrt{\frac{3}{2}}\Tilde{V}_{e^+}) & \frac{1}{3} & \frac{1}{3}(\Tilde{V}_{e^+}\Tilde{V}_{a\mu}\pm \sqrt{\frac{3}{2}}V_{a\tau})\\
        \frac{1}{3}(\Tilde{V}_{a\tau}\Tilde{V}_{e^+}\pm \sqrt{\frac{3}{2}}V_{a\mu}) & \frac{1}{3}(V_{a\mu}V_{e^+}\pm \sqrt{\frac{3}{2}}\Tilde{V}_{a\tau}) & \frac{1}{3}
    \end{bmatrix}
\end{equation}

\bigskip

\bigskip

\centerline{REFERENCES}

\bibliography{BMV}

\end{document}